\documentclass[11pt,a4paper]{JHEP3}
\usepackage{amsmath, amssymb}

\usepackage{epsfig,bbm,amssymb,euscript}
\usepackage{euscript}
\def\be{\begin{eqnarray}}
\def\ee{\end{eqnarray}}
\def\0{\nonumber}

\usepackage{amsfonts}
\usepackage{verbatim}
\newcommand\N{{\cal N}}

\newcommand\z{\zeta}

\newcommand\no{\noindent}

\def\k{\kappa}

\def\N{{\cal N}}

\def\sinh{{\rm sinh}}

\def\k{\kappa}

\def\bra#1{\langle #1 |}
\def\ket#1{|#1 \rangle}
\def\0{\nonumber}

\preprint{SISSA/45/2009/EP\\ ULB-TH/09-25\\\tt hep-th/0908.0056}

\title{Ghost story. III. Back to ghost number zero}

\author{ L.Bonora\\
International School for Advanced Studies (SISSA/ISAS)\\
Via Beirut 2--4, 34014 Trieste, Italy, and INFN, Sezione di
Trieste\\
E-mail:   \email{bonora@sissa.it},}

\author{C.Maccaferri\\
International Solvay Institutes and
Physique Th\'eorique et Math\'ematique,\\
ULB C.P. 231, Universit\'e Libre
de Bruxelles,  B-1050, Bruxelles, Belgium\\
E-mail:  \email{cmaccafe@ulb.ac.be},}

\author{D.D.Tolla\\
Department of Physics and University College,
Sungkyunkwan University,
Suwon 440-746, South Korea\\
E-mail:  \email{ddtolla@skku.edu}}

\abstract{After having defined a 3--strings midpoint--inserted vertex for the $bc$ system, we analyze the relation between $gh$=0 states (wedge states) and
$gh$=3 midpoint duals. We find explicit and regular relations connecting the two objects. In the case of wedge states this allows us to write down
a spectral decomposition for the $gh$=0 Neumann matrices, despite the fact that they are not commuting with the matrix representation of $K_1$. We thus
trace back the origin of this noncommutativity to be a consequence of the imaginary poles of the wedge eigenvalues in the complex $\k$--plane.
With explicit reconstruction formulas at hand for both $gh$=0 and $gh$=3, we can finally show how the midpoint vertex  avoids this intrinsic
noncommutativity at $gh$=0, making everything as simple as the zero momentum matter sector.}


\begin{document}

\maketitle
%%%%%%%%%%%%%%%%%%%%%%%%%%%%%%%%%%%%%%%%%%%%%%%%%%%%%%%
\section{Introduction}
%%%%%%%%%%%%%%%%%%%%%%%%%%%%%%%%%%%%%%%%%%%%%%%%%%%%%%%

This is the third and conclusive paper of a series including \cite{BMST}, referred to as I, and \cite{BMST2}, referred to as II; see also \cite{BST}.
The whole story started up with a very simple question.
In Open String Field Theory, \cite{Witten:1985cc}, wedge states are surface states $\ket N$
labeled by a real (possibly integer) number $N$, obeying the star algebra relation
$$\ket N*\ket M=\ket{N+M-1}.$$
Wedge states, originally defined and studied in \cite{Rastelli:2000iu, Rastelli:2001vb, Schnabl:2002gg}, have been a fundamental ingredient for constructing and analyzing, following the breakthrough by Schnabl \cite{Schnabl05}, analytic solutions
for the tachyon vacuum \cite{Okawa1, Fuchs0, Ellwood, RZ06, ORZ, Erler:2009uj, Erler:2007xt, Aref'eva:2008ad}, solutions describing
marginal deformations \cite{Okawa2, Okawa3, Schnabl:2007az, KORZ, Fuchs3, Lee:2007ns, Kwon:2008ap, Kiermaier:2007ki, Kiermaier:2007vu, Erler:2007rh} and
 related topics \cite{Fuchs1, Kroyter:2009zj, Kroyter:2009bg, Ellwood:2009zf, Kiermaier:2008qu, Ellwood:2008jh}. See \cite{Fuchs4} for a recent review
 and a guide through other references.\\

\no While this simple commutative $*$--subalgebra has a very intuitive realization in terms of gluing vertical strips in the
 $\arctan$--sliver frame (or, equivalently, of gluing sectors of cones), the oscillator realization
of this multiplication rule has only been understood in detail in the (non--universal) matter sector, \cite{Furu, Kishimoto}
 (corresponding to $D$--free bosons) and in the
(universal) bosonized ghost system, \cite{Belov1}. Strangely enough, the original (universal) $bc$ system which naturally emerges from the Polyakov path
integral, while quite similar to the matter sector in the conventional normal ordering on $c_1\ket0$, \cite{GJ1, GJ2, CST, Ohta, Samu},
turns out to give a quite complicated oscillator formalism, when one normal--orders with respect to the $SL(2,R)$--vacuum. This is due to several conspiring accidents but the very reason is the presence of the
three zero modes of the $c$ ghost. Indeed, a closely related system, the $h=(1,0)$ $bc$ system, having just one $c$ zero mode, turns out to be almost
identical to the zero momentum matter sector, \cite{MM, Kling:2003sb}.\\

In this paper we deal with the co--existence of the $gh$=0 and the $gh$=3 sectors. These two sectors are naturally conjugate to each other  by the
$bpz$ inner product. Still, while the $gh$=0 sector is a $*$--subalgebra, the $gh$=3 is not because the ghost number is additive under the $*$--product.
A related issue is that, in general, there is no unique prescription to pair $gh$=0 and $gh$=3 states
in a one--to--one way. Any insertion of 3 $c$'s operators could in principle do the job. There are however better choices than others.
In particular one would like to pair  $gh=0$ and $gh=3$ states in such a way that they retain the same conformal properties. This implies the use of a
$gh$=3 insertion with zero conformal weight. In the ghost sector, there is just one such field,  given by
$$
Y(z)=\frac12\partial^2c\partial c c(z).
$$
This composite field, also considered in \cite{GRSZ2} in the context of VSFT \cite{RSZ2},  has a number of nice properties
\begin{itemize}
\item It is a weight zero primary
$$T(z)Y(w)=\frac{\partial Y(w)}{z-w}+...\;\;.$$
\item It is BRST invariant
$$\{Q_B,Y(z)\}=\oint_z\, \frac{dw}{2\pi i}\, j_B(w) Y(z)=0.$$
\item Its derivative is $Q_B$--exact
$$\partial Y(z)=[Q_B,\partial^2 c\partial c(z)].$$ This implies that, given two BRST invariant $gh$=0 states (for example surface states) we have
$$\partial_z\bra{\psi_1}Y(z)\ket{\psi_2}=0,$$
So, for BRST--closed states, correlators are independent of the location of $Y$.
\end{itemize}
However OSFT deals with {\it off shell} states which need not be BRST--invariant. Moreover the $*$--product breaks  conformal
invariance down to the subgroup of midpoint preserving reparametrizations. This implies
that, out of the infinite places where one could insert Y, the only one which is not going to interfere is the midpoint. In general, when
dealing with conformal maps such as the ones which define the $*$--product, having midpoint insertions is forbidden because such maps are singular
precisely at the midpoint. The only exception is when we are dealing with weight zero primaries, which will not pick up any singular Jacobian from
the conformal transformation. Once again, at $gh$=3, this uniquely selects the operator $Y$. Concretely, the midpoint insertion of $Y$ commutes with
the generators of midpoint preserving reparametrizations
$$
[K_n,Y(\pm i)]=0.
$$
Here we will be mostly interested in $K_1$--invariance, since this symmetry allowed, in all previously known cases,
\cite{RSZ, Belov1, Belov2} to diagonalize the Neumann matrices
of the 3--string vertex.\\

\no Given the above reasons, for any state at $gh$=0, $\psi$, we define its midpoint dual(s) to be
$$
\hat \psi_{(\pm i)}\equiv Y(\pm i)\psi.
$$

\no This relation is not a one to one map in general because of the non trivial kernel of $Y(i)$ but it is
effectively so for $gh$=0 states which do not have $c$ insertions
at the midpoint. All $gh=0$ squeezed states (and in particular surface states) are in this class.\\

\no In this paper we deal with $gh$=0 and $gh$=3 squeezed states, where the latter are intended to be obtained from the former by the insertion of $Y$ at
one of the two midpoints $z=\pm i$. By using the method of images (doubling trick)  one can associate {\it two} $gh$=3 states to each $gh$=0 state
$$
\hat \psi\equiv Y\left(\frac\pi2\right)\psi=\frac12(Y(i)+Y(-i))\psi=\frac12(\hat \psi_{(i)}\oplus\hat \psi_{(- i)})\simeq(\hat \psi_{(i)},\,\hat \psi_{(-i)}).
$$
The reason for the above notation is that, under the $*$--product, squeezed states are a ring rather than an algebra.
This just means that we will separately consider states with a holomorphic and an anti--holomorphic midpoint insertion, because their half sum is
not a squeezed state anymore.\\

\no One of the main problems we address in this paper is how to get back, in the case of squeezed states (and, in particular, of surface states),
 the $gh$=0 states by knowing
their $gh$=3 cousins. To be more specific the $Y$ midpoint insertion will be implemented via a 2--strings vertex (midpoint reflector) which is
a squeezed state of total ghost number 6. The ordinary 2--strings vertex (implementing the $bpz$ conjugation) would have total ghost number 3. So the two differ
by 3 units of ghost number, which are provided by the $Y$ insertion.\\

\no We will see that $gh$=3 squeezed states which are obtained by applying $Y(\pm i)$ to $gh$=0 squeezed states, are characterized by having a Neumann matrix
whose zero mode part is uniquely determined by the non zero mode part. We call this property {\it midpoint identity}.\\
\no The explicit knowledge of $both$ $Y(\pm i)$--inserted $gh$=3 squeezed states, will allow us to reconstruct the original $gh$=0 squeezed state.
We will provide evidence that this is possible at least  for all (non--projectors) surface states.

\no After analyzing in detail the $gh$=0/$gh$=3 relations in the case of wedge states (which are the only $K_1$ invariant surface states), we
will finally $*$--multiply them via the midpoint 3-string vertex defined in paper II. Here we will see how the midpoint identities of the vertex,
together with the structure of the $gh$=0 Neumann matrices (which represent the states to be multiplied) conspire to effectively transform
these $gh$=0 matrices into $gh$=3 ones. Differently from $gh$=0 case, $K_1$--invariance at $gh$=3 does imply commuting matrices, so the star product will be
as easy as the zero momentum matter product. In particular all the matrices in the game will commute, so,  trading them with their eigenvalues,
they will reconstruct the star product of the two wedges with an overall $Y(\pm i)$ insertion. This insertion can then be undone by using
once again the $gh$=0/$gh$=3 relation and to finally get the $gh$=0 state which is just the star product of the 2 $gh$=0 states. This completes the program started with I.
%%%%%%%%%%%%%%%%%%%%%%%%%%%%%%%%%%%%%%%%%%%%%%%%%%%%%%%%
\section{Overview}
%%%%%%%%%%%%%%%%%%%%%%%%%%%%%%%%%%%%%%%%%%%%%%%%%%%%%%%%
The aim of this section is to give an overview of the objects that will feature in this paper.

\no We will deal with $gh$=0 states, the wedge states, and $gh$=3 states, wedge states with a midpoint $Y$ insertion. Both of them can be uniquely defined as squeezed states on
the vacua $\ket0$ and $\ket{\hat0}\equiv c_{-1}c_0c_1\ket0$, via a $gh$=0 Neumann function and a $gh$=3 one, respectively. The two different Neumann functions
are known once the conformal map is given. In the case of wedges, we fix $SL(2,C)$ invariance by choosing the conformal maps to be
\be
f_N(z)=\left(\frac{1+iz}{1-iz}\right)^{\frac2N}
\ee

%%%%%%%%%%%%%%%%%%%%%%%%%%%%%%%%%%%%%
\subsection{Wedge States}
%%%%%%%%%%%%%%%%%%%%%%%%%%%%%%%%%%%%%

 Wedge states are squeezed states on the $gh$=0 vacuum $\ket0$
\be
\ket {N}= e^{c_P^\dagger\,S_{Pq}\,b_q^\dagger}\ket0,
\ee
with the (long-short) defining matrix given by
\be\label{wedge-function}
S^{(N)}_{Mn}&=&\oint_0 \frac{dz}{2\pi i}\oint_0 \frac{dw}{2\pi i} \frac1{z^{M-1}}\frac1{w^{n+2}}\0\\
        &&\left[\frac{f_N'(z)^2}{f_N'(w)}\frac{1}{f_N(z)-f_N(w)}\left(\frac{f_N(w)-f_N(0)}{f_N(z)-f_N(0)}\right)^3-\frac{w^3}{z^3(z-w)}\right].
\ee
($M=-1,0,1,\ldots$, $n=2,3,\ldots$, see II for notation; moreover we will drop the label $N$ in $S^{(N)}$ whenever this is not strictly necessary).
This matrix has the general bulk/zero mode block decomposition $$S=\left(\begin{matrix}0&s\\ 0& S\end{matrix}\right),$$
where the upper left $0$ represents a vanishing $3\times3$ matrix, the lower left $0$ represents three infinite short columns,
$s$ represents three infinite short rows, while
$S$ is the bulk.
Wedge states are annihilated by $K_1=L_1+L_{-1}$. This can be checked explicitly by writing $K_1$ as
\be
K_1=  c_{N}^\dagger\, G_{NM} \,b_M +
b_{ n}^\dagger\,H_{nm}\,
c_{m}- 3c_2\,b_{-1}
\ee
We get $K_1\ket{N}=0$ iff
\be
(GS+SH^T)_{Nm}+3S_{N2}S_{-1m}=0
\ee
This relation can be explicitly  checked by picking up residues in eq.(\ref{wedge-function}).\\

\noindent Note that, despite the fact that $K_1\ket{N}=0$, the Neumann coefficients do not anti--commute with $G$ and $H$. Moreover the violation
of anti--commutativity depends explicitly on the zero mode contribution of the $gh$=0 Neumann function. This is one of the main reasons why it will be
necessary to insert the operator $Y$ at the midpoint and to consider $gh$=3 squeezed states, whose Neumann matrices will commute with $G$.
We will see later on that the knowledge of $gh$=3 midpoint inserted squeezed states will imply the knowledge of the $gh$=0 ones, whose $Y(\pm i)$ inserted versions give the former.

%%%%%%%%%%%%%%%%%%%%%%%%%%%%%%%%%%%%%%%%%%%%%%%%%%
\subsection{$Y(\pm i)$--inserted wedge states}
%%%%%%%%%%%%%%%%%%%%%%%%%%%%%%%%%%%%%%%%%%%%%%%%%%

We can define wedge states (and surface states in general) with local operator insertions. In particular we are interested in the midpoint insertion
of the BRST invariant, primary scalar $Y(z)=\frac12\partial^2c\partial c c(z)$.\\

Let us begin by  observing that a surface state (which is identified via an analytic map, $f(z)$,
from the unit semidisk of the UHP, to a Riemann surface $\Sigma$ with the disk topology)  with an insertion of $Y$ at a point $\xi$ on the Riemann
surface can be written as a squeezed state on the $gh$=3 vacuum $\bra{\hat0}=\bra0 c_{-1}c_0c_1$ as \footnote{Note that there is no normalization
arising from the transformation of the insertion, this is unambiguous because $Y$ is a zero--weight primary.}
\be
\bra{\hat S^\xi}=\bra{\hat0}\,e^{-c_n\,\hat S^{(\xi)}_{nM}\,b_M}
\ee
where the (short--long) Neumann coefficients are given by

\be\label{s3}
\hat S^{(\xi)}_{nM}&=&\oint_0 \frac{dz}{2\pi i}\oint_0 \frac{dw}{2\pi i} \frac1{z^{n-1}}\frac1{w^{M+2}}
        \left[\frac{f'(z)^2}{f'(w)}\frac{1}{f(z)-f(w)}\left(\frac{f(w)-\xi}{f(z)-\xi}\right)^3-\frac{1}{z-w}\right]
\ee

A midpoint insertion (which is only acceptable for zero--weight primaries) is given by $\xi=f(\pm i)$. \\

Let us concentrate on wedge states $\ket N$ whose analytic map (up to $SL(2,C)$) is given by $f_{N}(z)=\left(\frac{1+iz}{1-iz}\right)^{\frac2N}$. We have
respectively
\be
\bra{N}Y(i)&\equiv&\bra{\hat N_{(i)}}=\bra{\hat0}\,e^{-c_n\,\hat S^{(f_N(i))}_{nM}\,b_M}\\
\bra{N}Y(-i)&\equiv&\bra{\hat N_{(-i)}}=\bra{\hat0}\,e^{-c_n\,\hat S^{(f_N(-i))}_{nM}\,b_M},
\ee
In order to lighten a bit the notation we will rename
\be
\hat S_{(\pm i)}\equiv\hat S^{(f_N(\pm i))},
\ee
and write the Neumann coefficients as ($f_N(i)=0$, $f_N(-i)=\infty$)
\be
 {\hat S_{(i)}}{}_{nM}\!&=&\!\oint_0 \frac{dz}{2\pi i}\oint_0 \frac{dw}{2\pi i} \frac1{z^{n-1}}\frac1{w^{M+2}}
       \! \left[\frac{f_N'(z)^2}{f_N'(w)}\frac{1}{f_N(z)-f_N(w)}\!\left(\frac{f_N(w)}{f_N(z)}\right)^3\!-\!\frac{1}{z-w}\right]\label{Si}\\
{\hat S_{(-i)}}{}_{nM}\!&=&\!\oint_0 \frac{dz}{2\pi i}\oint_0 \frac{dw}{2\pi i} \frac1{z^{n-1}}\frac1{w^{M+2}}
        \left[\frac{f_N'(z)^2}{f_N'(w)}\frac{1}{f_N(z)-f_N(w)}-\frac{1}{z-w}\right]\label{S-i}
\ee
It is easy to prove the twist property
\be
{\hat S_{(i)}}{}_{nM}=\left({\hat S_{(-i)}}{}_{nM}\right)^*=(-1)^{m+N}{\hat S_{(-i)}}{}_{nM}.
\ee
From the fact that $$[K_1,Y(\pm i)]=0$$ it follows that $Y(\pm i)$--inserted  wedge states are also annihilated by $K_1=L_1+L_{-1}$.
This can be checked explicitly by writing $K_1$ as
\be
K_1=  c_{n}^\dagger\, H^T_{nm} \,b_m +
b_{ N}^\dagger\,G^T_{NM}\,
c_{M}+3c_{2}^\dagger\,b_{1}^\dagger.
\ee
We get $\bra{\hat N_{(\pm i)}}K_1=0$ iff
\be
(H^T\hat S_{(\pm i)}+\hat S_{(\pm i)} G)_{nM}+3\delta_{n2}\delta_{-1M}=0,
\ee
and it is easy to check this directly (see II).\\

\noindent We see  that the  Neumann coefficients still do not (anti--)commute with $G$ and $H$. The violation is however very mild
and, contrary to the $gh$=0 case, it is universal (it doesn't depend on the particular wedge state).

\noindent As argued in II, these $gh$=3 matrices can be `augmented' to auxiliary big matrices by adding the `radial ordering' matrix
$$z_{ij}=\delta_{i+j},\quad\quad i,j=-1,0,1\quad,$$
to the 3x3 zero mode block, in the following way
\be
{\bf \hat S}_{(\pm i)}=\left(\begin{matrix}z&\quad 0\\ \hat s_{(\pm i)}&\quad \hat S_{(\pm i)}\end{matrix}\right).
\ee
This extra term will not contribute to the state, because of normal ordering, and it is thus a freedom we can take.
It is then immediate to check that
\be
G{\bf \hat S}_{(\pm i)}+{\bf \hat S}_{(\pm i)}G=0
\ee

%%%%%%%%%%%%%%%%%%%%%%%%%%%%%%%%%%%%%%%%%%%%%%%%%
\subsubsection{Principal $gh$=3 wedges}
%%%%%%%%%%%%%%%%%%%%%%%%%%%%%%%%%%%%%%%%%%%%%%%%%

In order to understand the spectral decomposition of the above defined Neumann matrices (both at $gh$=0 and at $gh$=3),
it will be useful to define auxiliary twist invariant $gh$=3 squeezed states, which we dubbed in II  `principal' $gh$=3 wedges.
Principal  wedge states are twist invariant squeezed states on the $gh$=3 vacuum $\bra{\hat0}\equiv \bra0c_{-1}c_0c_1$
\be
\bra {\hat N}= \bra{\hat0}\,e^{-c_n\,\hat S_{nM}\,b_M},
\ee
with the (short-long) defining matrix given by\footnote{This expression is not $SL(2,C)$ invariant. This is the first indication that these states are not
surface states with insertions.}
\be
\hat S_{pQ}&=&\oint_0\frac{dz}{2\pi i}\oint_0\frac{dw}{2\pi i}\frac{1}{z^{p-1}}\frac{1}{w^{Q+2}}\,
\left(\frac{2i}{N}\,\frac{1+w^2}{(1+z^2)^2}\,\frac{f_N(z)+f_N(w)}{f_N(z)-f_N(w)}-\frac1{z-w}\right)\\
\ee
This matrix has the general form $$\hat S=\left(\begin{matrix}0&0\\ \hat s& \hat S\end{matrix}\right).$$\\

We will see in the next section that these states differ from wedges with $Y(\pm i)$ insertion by a rank 2 matrix. It follows
that they are not surface states with insertions and, consequently,
they are not BRST invariant (see appendix A for a discussion concerning BRST invariance and the overlap between $gh$=0 and $gh$=3 squeezed states).\\
\noindent Their Neumann matrix is nonetheless the `principal' part of all the $gh$=0 and $gh$=3 BRST invariant Neumann matrices. {\it All other
matrices will differ from the former by a rank 2 correction}. Principal $gh$=3 wedge states are also annihilated by $K_1=L_1+L_{-1}$
and, as we saw in paper II, their bulk is completely reconstructed from the continuous spectrum,
with the standard choice of contour given by the  real axis of the complex $\k$--plane.

%%%%%%%%%%%%%%%%%%%%%%%%%%%%%%%%%%%%%%%%%%%%%%%%%%%%%%%%%%%%%%%%%%%%%%%%%%
\section{Reconstruction formulas for $gh$=3 wedges}
%%%%%%%%%%%%%%%%%%%%%%%%%%%%%%%%%%%%%%%%%%%%%%%%%%%%%%%%%%%%%%%%%%%%%%%%%%

Since the $gh$=3 Neumann matrices (anti)--commute with $G$, the standard continuous plus discrete basis will
reconstruct them starting from their eigenvalues.
For the sake of clarity, here we briefly review some of the results derived in paper II.

%%%%%%%%%%%%%%%%%%%%%%%%%%%%%%%%%%%%%%%%%%%%%%%%
\subsection{Principal $gh$=3 wedges}
%%%%%%%%%%%%%%%%%%%%%%%%%%%%%%%%%%%%%%%%%%%%%%%%

These are the simplest states, as their bulk is just given by the real continuous spectrum,
(and it is the `principal part' for all the $gh$=3/0 states we will
consider)

\be
\hat S_{pq}&=&\int d\k\, \frac1{2\sinh\frac {\pi \k}{2}}\,\frac{\sinh\frac{\pi \k (2-N)}{4}}{\sinh\frac{\pi \k N}{4}}V_p^{(2)}(-\k)V_q^{(-1)}(\k),
\quad\quad p,q\geq2.
\ee
The complete reconstruction, including the zero modes 3--column and the radial ordering block $z_{ij}=\delta_{i+j}$,
is given by the continuous + discrete spectrum
\be
\hat S_{PQ}&=&\int d\k\, \frac1{2\sinh\frac {\pi \k}{2}}\,\frac{\sinh\frac{\pi \k (2-N)}{4}}{\sinh\frac{\pi \k N}{4}}V_p^{(2)}(-\k)V_Q^{(-1)}(\k)\0\\
&+&\sum_{\xi=\pm 2i,0}  (-1)^{\frac\xi{2i}}\tilde v_P^{(2)} (-\xi)  \tilde V_Q^{(-1)}(\xi)\0\\
&=&\left(\begin{matrix}z_{ij}&\quad 0\\ \hat s_{pi}&\quad \hat S_{pq}\end{matrix}\right).
\ee
We recall, from paper II, that  the $\tilde v^{(2)}(\xi)$ discrete eigenvectors are given by
\be
\tilde v^{(2)}(0)= \frac1{\sqrt{2}}\left(\begin{matrix} 1\\0\\1\\0\\-3\\0\\ \dot :\end{matrix}\right),
\quad\quad
\tilde v^{(2)}(\pm 2i)= \frac12\left(\begin{matrix} 1\\\mp i\\-1\\ \pm i\\ 1\\ \mp i\\
\dot :\end{matrix}\right),\label{V(2)}
\ee
while the $\tilde V_Q^{(-1)}(\xi)$ are just the $h=-1$ continuous eigenvectors, evaluated at $\xi=0,\pm 2i$, with the same normalization as the $\tilde v^{(2)}(\xi)$'s.
\be
\tilde V^{(-1)}(0)&=&\frac1{\sqrt{2}}\,V^{(-1)}(0)=\frac1{\sqrt{2}}\,(1,0,1;\;0,0,0,...)\\
\tilde V^{(-1)}(\pm 2i)&=&\frac1{2}\,V^{(-1)}(\pm 2i)=\frac1{2}\,(1,\pm 2i,-1;\;0,0,0,...).
\ee
These vectors just contain zero mode contributions. Moreover the continuous spectral measure
for the identity operator, $\frac 1{2\sinh\frac{\pi\k}{2}}$, has simple poles precisely at
$(\k=0,\pm 2i)$. This can be viewed as the origin of the discrete spectrum (for a
thorough discussion see II).

%%%%%%%%%%%%%%%%%%%%%%%%%%%%%%%%%%%%%%%%%%%%%%%%%%%%%%%%%
\subsection{$Y(\pm i)$--inserted $gh$=3 wedges}
%%%%%%%%%%%%%%%%%%%%%%%%%%%%%%%%%%%%%%%%%%%%%%%%%%%%%%%%%

The Neumann functions of the  $gh$=3 wedges $\bra{\hat N_{(\pm i)}}\equiv\bra N Y(\pm i)$, and the one of the `principal' wedges $\bra{\hat N}$, are  very simply related. It is a
matter of trivial algebra to check that for $\bra{\hat N_{(i)}}$ we have
\be
\hat S_{(i)}(z,w)&=&\frac{f_N'(z)^2}{f_N'(w)}\frac{1}{f_N(z)-f_N(w)}\left(\frac{f_N(w)}{f_N(z)}\right)^3\0\\
&=&\frac{2i}{N}\,\frac{1+w^2}{(1+z^2)^2}\,\frac{f_N(z)+f_N(w)}{f_N(z)-f_N(w)}-\frac{4i}{N}\frac{1+w^2}{(1+z^2)^2}
\left(\frac{f_N(w)}{f_N(z)}+\frac12\right),
\ee
while for $\bra{\hat N_{(- i)}}$
\be
\hat S_{(-i)}(z,w)&=&\frac{f_N'(z)^2}{f_N'(w)}\frac{1}{f_N(z)-f_N(w)}\0\\
 &=&\frac{2i}{N}\,\frac{1+w^2}{(1+z^2)^2}\,\frac{f_N(z)+f_N(w)}{f_N(z)-f_N(w)}+
\frac{4i}{N}\frac{1+w^2}{(1+z^2)^2}\left(\frac{f_N(z)}{f_N(w)}+\frac12\right).
\ee
This readily gives the following reconstruction formulas
\be
\hat S_{(i)}(z,w)&=&\hat S(z,w)-\frac{4i}{N}\left(f^{(2)}_{\k=-\frac{4i}{N}}(z)f^{(-1)}_{\k=\frac{4i}{N}}(w)+\frac12f^{(2)}_{\k=0}(z)f^{(-1)}_{\k=0}(w)\right)\\
\hat S_{(-i)}(z,w)&=&\hat S(z,w)+\frac{4i}{N}\left(f^{(2)}_{\k=\frac{4i}{N}}(z)f^{(-1)}_{\k=-\frac{4i}{N}}(w)+\frac12f^{(2)}_{\k=0}(z)f^{(-1)}_{\k=0}(w)\right),
\ee
where $\hat S(z,w)$ is the Neumann function for the `principal' $gh$=3 wedges (whose bulk is just given by the continuous spectrum) and
\be
f^{(2)}_{\k}(z)&=&\frac1{(1+z^2)^2}\,e^{\k\tan^{-1}z}=\sum_{n=2}^\infty\,V_n^{(2)}(\k)\,z^{-n+1}\0\\
f^{(-1)}_{\k}(z)&=&(1+z^2)\,e^{\k\tan^{-1}z}=\sum_{M=-1}^\infty\,V_M^{(-1)}(\k)\,z^{-M-2}.\0
\ee
It is easy to see that the above reconstruction formulas can (almost) be obtained
by integrating (we focus on $\hat S_{(i)}(z,w)$)  on a path along the band $\frac{4}{N}<\Im(\k)<2$. This gives (in addition to the contribution along the
real axis) the residue around $\k=\frac{4i}{N}$ and {\it half} the residue at $\k=0$ (this pole counts $\frac12$ because it sits on the real axis).
Concretely
\be
&&\int_{\frac{4}{N}<\Im(\k)<2} d\k\, \frac1{2\sinh\frac {\pi \k}{2}}\,\frac{\sinh\frac{\pi \k (2-N)}{4}}{\sinh\frac{\pi \k N}{4}}f_{-\k}^{(2)}(z)f_\k^{(-1)}(w)=
\int_R d\k\, (...)
-\oint_{\frac{4i}{N}}d\k(...)-\frac12\oint_0d\k(...)\0\\
&&=\hat S(z,w)|_{cont}-\frac{4i}{N}f^{(2)}_{\k=-\frac{4i}{N}}(z)f^{(-1)}_{\k=\frac{4i}{N}}(w)-
\frac{i(2-N)}{N}\,f^{(2)}_{\k=0}(z)f^{(-1)}_{\k=0}(w)\0\\
&&=\hat S_{(i)}(z,w)|_{cont}+if^{(2)}_{\k=0}(z)f^{(-1)}_{\k=0}(w),
\ee
where by the subscript $(\cdot)|_{cont}$, we indicate the contribution from just the continuous spectrum.\\

\noindent One should notice that a  small mismatch is present: the half residue around $\k=0$  gives an extra
$i f^{(2)}_{\k=0}(z)f^{(-1)}_{\k=0}(w)$. It should be stressed however that this extra contribution
is $\it universal$, i.e. it does not depend on the $N$ of the wedge state under consideration, moreover it only affects the  3--column of the Neumann
matrix. We will see in a while that this extra universal contribution is not an unwelcome
accident but a manifestation of a very important relation: the midpoint identity.\\

\noindent In conclusion, modulo $\it universal$ stuff, $Y(i)$--inserted $gh$=3 wedges are precisely reconstructed by shifting
the path of the $\k$ integration.  \\

\noindent To this one should add the contribution from the discrete spectrum, which is the same as for `principal' states. In total we thus have
 ($\xi_N=+\frac{4i}{N}$)
\be
\hat S_{(i)}{}_{MQ}&=&\int_{\Im(\k)>\Im(\xi_N)} d\k\, \frac1{2\sinh\frac {\pi \k}{2}}\,\frac{\sinh\frac{\pi \k (2-N)}{4}}{\sinh\frac{\pi \k N}{4}}V_m^{(2)}(-\k)V_Q^{(-1)}(\k)\0\\
&+&\sum_{\xi=\pm 2i,0}  (-1)^{\frac\xi{2i}}\tilde v_M^{(2)} (-\xi)  \tilde V_Q^{(-1)}(\xi)\0\\&-&iP_{mQ}\0\\
&=&\left(\begin{matrix}z_{ij}&\quad 0\\ \hat s_{(i)}{}_{pj}&\quad \hat S_{(i)}{}_{mq}\end{matrix}\right),
\ee
where, following II\footnote{In II the states augmented with $z$ were denoted by a prime, $S'$.
Now we think we can drop the prime without harm.}, we have indicated
\be
P_{mQ}=V_m^{(2)}(0)V_Q^{(-1)}(0).
\ee
As anticipated above, the spurious $-iP_{mQ}$ contribution will find its rationale in the implementation of the midpoint identity, as we will see in the
next section.\\

The same reconstruction obviously apply to $S_{(-i)}(z,w)$,
by just integrating along the band $-2<\Im(\k)<-\frac{4}{N}$. Explicitly
\be
\hat S_{(-i)}{}_{MQ}&=&\int_{\Im(\k)<-\Im(\xi_N)} d\k\, \frac1{2\sinh\frac {\pi \k}{2}}\,\frac{\sinh\frac{\pi \k (2-N)}{4}}{\sinh\frac{\pi \k N}{4}}V_m^{(2)}(-\k)V_Q^{(-1)}(\k)\0\\
&+&\sum_{\xi=\pm 2i,0}  (-1)^{\frac\xi{2i}}\tilde v_M^{(2)} (-\xi)  \tilde V_Q^{(-1)}(\xi)\0\\&+&iP_{mQ}\0\\
&=&\left(\begin{matrix}z_{ij}&\quad 0\\ \hat s_{(-i)}{}_{pj}&\quad \hat S_{(-i)}{}_{mq}\end{matrix}\right).
\ee

%%%%%%%%%%%%%%%%%%%%%%%%%%%%%%%%%%%%%%%%%%%%%%%%%%%%%%%%%%
\section{ The relation between $gh$=3 and $gh$=0 }
%%%%%%%%%%%%%%%%%%%%%%%%%%%%%%%%%%%%%%%%%%%%%%%%%%%%%%%%%%

In the last section we gave the reconstruction formulas for $gh$=3 wedge states. We saw at the beginning that it is not going to be straightforward
to write down a corresponding formula for $gh$=0 states, the reason
being that $gh$=0 Neumann matrices do not (anti)--commute with $G$, so the
relevant reconstruction formula cannot involve simply the continuous and
discrete eigenvectors of $G$ (no matter whether $\k$ is complex, as for all values
of $\k$ we formally have eigenvectors of $G$).\\

%%%%%%%%%%%%%%%%%%%%%%%%%%%%%%%%%
\subsection{The midpoint reflector}
%%%%%%%%%%%%%%%%%%%%%%%%%%%%%%%%%

The strategy we are going to follow is thus a bottom up one and it is based on the midpoint 2--string vertex (reflector).  This vertex is
the oscillator incarnation of the insertion of $Y$ at the midpoint. Since we have `two' midpoints (the holomorphic and the anti--holomorphic one),
the midpoint insertion will be given by the following reflector(s)
\be
\bra{ V^{(2)}}(Y(i)\,,\,Y(-i))=\left(\bra{ V_{(+i)}^{(2)}}\,,\,\bra{ V_{(-i)}^{(2)}}\right)
\ee
where $\bra{ V^{(2)}}$ is just the usual 2--string vertex which implements bpz conjugation.

\noindent The midpoint reflector is meant to send a $gh$=0 right state (wedge state) to a $gh$=3 left state ($Y(i)$--inserted wedge) as follows
\be
\left(\bra{ V_{(+i)}^{(2)}}\,,\,\bra{ V_{(-i)}^{(2)}}\right)\,\ket N=\left(\bra{\hat N_{(+i)}}\,,\,\bra{\hat N_{(-i)}}\right)
\ee

\noindent Each of the two entries is a squeezed state  given by
\be
\bra{ V_{(\pm i)}^{(2)}}&=&\bra{ V^{(2)}}Y(\pm i)\0\\
&=& \,\!_{12}\!\bra{\hat0}\exp\left(-\sum_{n,M}\,c^r_n\,\hat R^{rs}_{(\pm i)}{}_{nM}\,b^s_M\right),\quad n\geq2,M\geq-1,\quad r,s=1,2
\ee
We concentrate from now on on the $Y(+i)$ insertion, as all the results for the $Y(-i)$ insertion can be
obtained by simple complex conjugation.
The Neumann coefficients are given by ($f_r(i)=f_s(i)$)

\be
\hat R^{rs}_{(i)}{}_{nM}&=&\oint_0 \frac{dz}{2\pi i}\oint_0 \frac{dw}{2\pi i} \frac1{z^{m-1}}\frac1{w^{N+2}}R_{(i)}^{rs}(z,w)\\
\hat R^{rs}_{(i)}(z,w)&=& \langle Y(f_{r,s}(i)) f_r\circ b(z) f_s\circ c(w)\rangle-\frac{\delta^{rs}}{z-w}\0\\
&=&\frac{f_r'(z)^2}{f_s'(w)}\frac{1}{f_r(z)-f_s(w)}\left(\frac{f_r(w)}{f_s(z)}\right)^3-\frac{\delta^{rs}}{z-w},
\ee
where the $SL(2,C)$ gluing functions are
\be
f_r(z)=(-1)^r\,\frac{1+iz}{1-iz}.
\ee
The `principal'/`residual' decomposition is given by
\be
\hat R^{rs}_{(i)}(z,w)&=&\hat R^{rs}(z,w)-\frac{4i}{2}\left((-1)^{r-s}f^{(2)}_{-\frac{4i}{2}}(z)f^{(-1)}_{\frac{4i}{2}}(w)+\frac12f^{(2)}_{\k=0}(z)f^{(-1)}_{\k=0}(w)\right),
\ee
and the `principal' part is
\be
\hat R^{rs}(z,w)=\left(\frac{2i}{2}\,\frac{1+w^2}{(1+z^2)^2}   \,\frac{f_r(z)+f_s(w)}{f_r(z)-f_s(w)}-\frac{\delta^{rs}}{z-w}\right).
\ee
As usual, in addition to the `principal' part we have `residual' contributions from
$\k=\pm\frac{4i}2=\pm2i$ and $\k=0$.
\no By explicitly computing the Neumann coefficients we get the following  matrices
\be
\hat R_{(i)}^{11}=\hat R_{(i)}^{22}\equiv \hat R_{(i)}\!&=&\!
\left(\begin{matrix}0&0\\r_{ni}&0\end{matrix}\right)=\left(\begin{matrix}0&0\\r&0\end{matrix}\right)\\
\hat R_{(i)}^{12}=\hat R_{(i)}^{21}\equiv \hat R'_{(i)}\!&=&\!
\left(\begin{matrix}0&0\\(-1)^{n+1}r^*_{n,-i}&(-1)^n\delta_{nm}\end{matrix}\right)=\left(\begin{matrix}0&0\\-{\cal C}r^*z&{\cal C}\end{matrix}\right)
\ee

\no The 3--column $r$ is just the Neumann matrix of the  $N=2$  wedge (vacuum) with the  $Y(+i)$ insertion, this matrix has no bulk but just three columns.
Explicitly
\be
r(z,w)&=&\sum_{n\geq2}\sum_{|i|\leq 1}\, r_{ni} z^{-n+1}w^{-i-2}=\frac1{z-w}\left(\frac{w-i}{z-i}\right)^3-\frac1{z-w}\\
&=&t(z,w)-2i\left(f^{(2)}_{-2i}(z)f^{(-1)}_{2i}(w)+\frac12f^{(2)}_{\k=0}(z)f^{(-1)}_{\k=0}(w)\right)\0,
\ee
Here $t(z,w)$ is the Neumann function of the `principal' $N=2$ $gh$=3 wedge. \\

\no We now consider the most general $gh$=0 squeezed state
\be
\ket S=e^{c^\dagger_N\,{\cal S}_{Nm}\,b^\dagger_m}\,\ket0\0,
\ee
the matrix $S$ has the structure
\be
{\cal S}=\left(\begin{matrix}0&s\\0&S\end{matrix}\right).
\ee
Now we reflect it by adapting the general method of \cite{KPot}
\be
\bra{\hat S_{(i)}}\equiv\bra{ V^{(2)}_{(i)}}\,\ket S=\,\det(1-{\cal S}\hat R_{(i)})\,
\bra{\hat0}\exp\left(-c_n\,\left[\hat R_{(i)}+\hat R'_{(i)}\frac1{1-{\cal S}\hat R_{(i)}}{\cal S}R'_{(i)}\right]_{nM}\,b_M\right)\0
\ee
Using the block decomposition, we have
\be
\frac1{1-{\cal S}\hat R_{(i)}}=\left(\begin{matrix}\frac1{1-sr}&| & 0\\-------&|&---- \\Sr\frac1{1-sr}&| &1 \end{matrix}\right),
\ee
and
\be
\det(1-{\cal S}\hat R_{(i)})=\det_{3\times3}\left(1-sr\right)
\ee
In total, after some trivial algebra, we  get
\be
\hat R_{(i)}+\hat R'_{(i)}\frac1{1-{\cal S}\hat R_{(i)}}{\cal S}\hat R'_{(i)}=
\left(\begin{matrix}0&\quad0\\ r-\hat S_{(i)}r^*z&\quad\hat S_{(i)}\end{matrix}\right)
\ee
\be\label{gh0/gh3}
\hat S_{(i)}&=&{\cal C}\left(S+(Sr-r^*z)\frac1{1-sr}\,s\right){\cal C}
\ee
A clarification is in order. We are comparing the Neumann matrix of a {\it ket} state, $S$, with the one of a $bra$, $\hat S_{(i)}$. As we are going to
check these relations explicitly for wedge states, this does not cause any problem, because at $gh$=0 we have ${\cal C}S{\cal C}=S$. However, in view of other possible applications,
we should write the above $gh$=0/$gh$=3 relation in an unambiguous way, even for ($gh=0$) states that are not twist invariant. In doing this we have to declare if
the squeezed state matrices are the one relative to the $bra$ or the $ket$ representation. Since surface states are usually defined as a $bra$, we reserve
the notation $S$ for the Neumann matrices of $bra$--states. So, with these conventions, the most general $ket$--squeezed state is given by
\be
\ket S=e^{c^\dagger_N\,[{\cal C}{\cal S}{\cal C}]_{Nm}\,b^\dagger_m}\,\ket0\0,
\ee
With this convention, using ${\cal C}r{\cal C}=r^*$, the equation (\ref{gh0/gh3}) becomes
\be\label{gh0/gh3'}
\hat S_{(i)}&=&\left(S+(Sr^*-rz)\frac1{1-sr^*}\,s\right).
\ee
All the matrices entering this equation are the ones which define $bra$ states, both at $gh=0$ and at $gh=3$.\\

\no The $Y(-i)$--inserted $gh$=3 Neumann matrix (which is obtained from the reflector $\bra{V^{(2)}_{(-i)}}$) will be thus

\be\label{gh0/gh3(-i)}
\hat S_{(-i)}&=&\left(S+(Sr-r^*z)\frac1{1-sr}\,s\right).
\ee

%%%%%%%%%%%%%%%%%%%%%%%%%%%%%%%%%%%%%%%%%%%%
\subsection{Midpoint Identities}
%%%%%%%%%%%%%%%%%%%%%%%%%%%%%%%%%%%%%%%%%%%%

The first thing to notice is that when we reflect a $gh$=0 state, the resulting $gh$=3 state will have the leftmost 3--column that is determined
 by the $gh$=3 bulk $\hat S_{(i)}$, that is
all the $gh$=3 squeezed states that we get by reflecting $gh$=0 squeezed states will have a defining matrix of the form
$$\left(\begin{matrix}0 & 0\\\hat s_{(i)}&\hat S_{(i)} \end{matrix}\right),$$ where the 3--column is given by
\be
\hat s_{(i)}=r-\hat S_{(i)}r^*z\label{mid-id}
\ee
Following the terminology of \cite{Erler:2004hv} (where a similar relation for the $gh$=1/$gh$=2 doublet was discussed, see also \cite{Oku2, tope})
we call this property {\it midpoint identity}.\\

\noindent Since we expect to get the $Y$--inserted $gh$=3 wedges by reflecting the $gh$=0 ones, the first consistency check is to see if the $gh$=3 Neumann functions
obey the midpoint identity. \\

\noindent The proof of $gh$=3 midpoint identities is a perfect playground to see the power of reconstruction formulas at work. \\
It is indeed sufficient to use the previously stated reconstructions,  plus the  orthogonality relation of the (-1,2) basis
\be\label{ortho}
\sum_{n\geq 2}V_n^{(-1)}(x)V_n^{(2)}(y)=2\sinh\frac{\pi x}{2}\delta(x-y).
\ee
This orthogonality condition is valid for general complex $(x,\; y)$, but when $x$ coincides with a pole in the measure $\frac{1}{2\sinh\frac{\pi x}{2}}$,
great care must be exercised.

%%%%%%%%%%%%%%%%%%%%%%%%%%%%%%%%%%%%%%%%%%%%%%%%%%%%%%%%%%
\subsubsection{Midpoint identities for `principal' states}
%%%%%%%%%%%%%%%%%%%%%%%%%%%%%%%%%%%%%%%%%%%%%%%%%%%%%%%%%%%%

We will first prove a related midpoint identity which links the 3--column and the bulk of the {\it principal} $gh$=3 wedges. This will involve the use
of the discrete and continuous spectrum, as reviewed above from II. The complex midpoint identity we want to prove will be easily obtained by just parallel--shifting
the path of the $\k$ integration at $\Im(\k)>2$. The `principal' midpoint identity is

\be\label{midzero}
(\hat S tz)_{pj}=(t-\hat s)_{pj}\,\quad\quad(z_{ij}=\delta_{i+j}), \quad -1\leq i,j \leq 1
\ee
The short--short matrix $\hat S$ is the bulk of the `principal' $gh$=3 wedge
\be
\hat S_{pq}&=&\int_{-\infty}^{\infty}\, d\k\, \frac1{2\sinh\frac {\pi \k}{2}}\,
\frac{\sinh\frac{\pi \k (2-N)}{4}}{\sinh\frac{\pi \k N}{4}}V_p^{(2)}(-\k)V_q^{(-1)}(\k)
\ee
The 3--column $\hat s_{pj}$ is the zero mode part of the  `principal' $gh$=3 wedge
(whose reconstruction is given by both the continuous and the discrete spectrum)
\be
\hat s_{pj}&=&\int_{-\infty}^{\infty} d\k\, \frac1{2\sinh\frac {\pi \k}{2}}\,
\frac{\sinh\frac{\pi \k (2-N)}{4}}{\sinh\frac{\pi \k N}{4}}V_p^{(2)}(-\k)V_j^{(-1)}(\k)\0\\
&+&\sum_{\xi=\pm 2i,0} \, (-1)^{\frac\xi{2i}}\,\tilde v_p^{(2)} (-\xi) \,  \tilde V_j^{(-1)}(\xi)
\ee
The 3--column $t_{pj}$ is the Neumann matrix of the $N=2$ `principal' $gh$=3 wedge. In turn the same quantity arises in the `lame' completeness relation of the continuous
and discrete spectrum in the zero mode sector. In particular
\be
 t_{nj}&=&-\int_{-\infty}^{\infty} d\k\, \frac1{2\sinh\frac {\pi \k}{2}}\,V_n^{(2)}(\k)V_{-j}^{(-1)}(\k)
=\sum_{\xi=\pm2i,0}\,\tilde v^{(2)}_{n}(\xi)\tilde V^{(-1)}_{-j}(\xi).
\ee
We remark that in II this quantity was denoted in a different way,
\be
 t_{nj}=-b_{-jn}.\0
\ee
A consequence of the above relations is the fact that the zero modes of the $h=-1$ basis are expressible as a linear combination of
the non--zero modes
\be
V_i^{(-1)}(\k)=-V_n^{(-1)}(\k)\,t_{n,-i}\;,\quad\quad \k\in R.
\ee
Using the above equations it is then immediate to prove (\ref{midzero}) in the form
\be
\hat s_{ni}=(t-\hat Stz)_{ni}\0
\ee

%%%%%%%%%%%%%%%%%%%%%%%%%%%%%%%%%%%%%%%%%%%%%%%%%%%%%%%%%%%%%%%%%
\subsubsection{Midpoint identities for $Y(i)$--inserted wedges}
%%%%%%%%%%%%%%%%%%%%%%%%%%%%%%%%%%%%%%%%%%%%%%%%%%%%%%%%%%%%%%%%%

Given the previous result for `principal' $gh$=3 wedges, we can now easily prove the midpoint identity for the $Y(i)$ insertion
$$\hat s_{(i)}=r-\hat S_{(i)}r^*z.$$

\no To this end it is important to notice that the 3--column $r^*_{ni}$ has {\it the same} integral representation as $t_{ni}$ but with a shifted path
in the complex $\k$ plane with $\Im(\k)>2$ \footnote{The same is true for $r_{ni}$ where the path is shifted at $\Im(\k)<-2$.}
\be
r^*_{ni}=-\int_{\Im(\k)>2}\,d\k\,\frac1{2\sinh\frac {\pi \k}{2}}\,V_n^{(2)}(\k)V_{-i}^{(-1)}(\k)
\ee
This implies that we have
\be
V_{i}^{(-1)}(\k)=-V_{n}^{(-1)}(\k)\,r^*_{n,-i}\,,\quad\quad 2<\Im(\k)<4
\ee
We can now easily compute $\hat S_ir^*z$ by using (\ref{ortho}) and we get
\be
[\hat S_{(i)}r^*z]_{nj}&=&\int_{\Im(\k)>\frac{4}{N}}\,d\k\,\frac{{\mathfrak t}_N(\k)}{2\sinh\frac {\pi \k}{2}}\,V_n^{(2)}(-\k)V_{m}^{(-1)}(\k)\0\\
&\times&\left(-\int_{\Im(\k')>2}\,d\k'\,\frac1{2\sinh\frac {\pi \k'}{2}}\,V_m^{(2)}(\k')V_{j}^{(-1)}(\k')\right)\0\\
&=&-\int_{\Im(\k')>2}\,d\k'\,\frac{{\mathfrak t}_N(\k')}{2\sinh\frac {\pi \k'}{2}}\,V_m^{(2)}(-\k')V_{j}^{(-1)}(\k')
\ee
It is important to notice that, to get the last line of the above equation, we have disregarded the secondary poles of
$${\mathfrak t}_N(\k)=\frac{\sinh\frac{\pi \k (2-N)}{4}}{\sinh\frac{\pi \k N}{4}},$$
at $\k=n\xi_N$ for $n>1$. Starting from the wedge $N=5$ these poles would begin to give contribution when the path is shifted from $\k>\xi_N$ to $\k>2i$.
The reason why they must be neglected is explained in appendix B, where the relevant part of this computation is performed in a regularized
way on the original $z$--plane.\\

\no In order to use the result we got for the `principal' part (`principal' states) we deform the path back to the real line and we pick up the residues
(from just the principal poles) along the way
\be
\int_{\Im(\k')>2}=\int_R\,-\oint_{2i}\,-\oint_{\frac{4i}{N}}\,-\frac12\oint_0\,,
\ee
which gives (using the midpoint identity for the `principal' states)
\be
[\hat S_{(i)}r^*z]_{nj}&=&t_{nj}-\hat s_{nj}-2iV_m^{(2)}(-2i)V_{j}^{(-1)}(2i)+\xi_N V_m^{(2)}(-\xi_N)V_{j}^{(-1)}(\xi_N)\0\\&&\!+
\left(\frac{\xi_N}{2}-i\right)V_m^{(2)}(0)V_{j}^{(-1)}(0)\0.
\ee
Remembering now
\be
\hat s_{(i)}{}_{nj}&=&\hat s_{nj}-\xi_N\left(V^{(2)}_n(-\xi_N)V^{(-1)}_j(\xi_N)+\frac12V^{(2)}_n(0)V^{(-1)}_j(0)\right)\\
r_{nj}&=&[\hat s_{(i)}|_{N=2}]_{nj}= t_{nj}-2i\left(V^{(2)}_n(-2i)V^{(-1)}_j(2i)+\frac12V^{(2)}_n(0)V^{(-1)}_j(0)\right),
\ee
we get finally
\be
[\hat S_{(i)}r^*z]_{nj}&=&[r-\hat s_i]_{nj}
\ee
Notice that the universal contribution at $\k=0$, $$-iP_{mj}=-iV_m^{(2)}(0)V_{j}^{(-1)}(0),$$ which comes from the wedge eigenvalue at $\k=0$, is there
precisely to implement the midpoint identity.

%%%%%%%%%%%%%%%%%%%%%%%%%%%%%%%%%%%%%%%%%%%%%%%%%%%%%%%%%%%%
\subsubsection{Midpoint identities for the vertex}
%%%%%%%%%%%%%%%%%%%%%%%%%%%%%%%%%%%%%%%%%%%%%%%%%%%%%%%%%%%%

For the 3--strings vertex defined in paper II ($Y(+i)$--insertion), the  midpoint identities easily generalizes to
\be
\hat v_{(i)}^{ab}&=&\delta^{ab}r-\hat V_{(i)}^{ab}r^*z,
\ee

\no To see this we have to compute, using the same strategy as before,
\be
[\hat V_{(i)}^{ab}r^*z]_{ni}=-\int_{\Im(\k)>2}d\k\,\frac{v^{ab}(\k)}{2\sinh\frac {\pi \k}{2}}\,V_n^{(2)}(-\k)V_{i}^{(-1)}(\k).\0
\ee
It is important to notice that (see paper II)
\be
v^{ab}(\k=\pm2i)&=&\delta^{ab},\quad\quad v^{ab}(\k=0)=\frac23-\delta^{ab}\0
\ee
Taking back the path to the real axis
$$
\int_{\Im(\k)>2}=\int_{R}-\oint_{2i}-\oint_{\frac{4i}{3}}-\frac12\oint_{0}\,,
$$
and using
\be
2\pi i\,\text{Res}_{\k=\frac{4i}{3}}\,\frac{v^{ab}(\k)}{2\sinh\frac {\pi \k}{2}}&=&\frac{4i}{3}\,\alpha^{a-b},\quad\quad \alpha=e^{\frac{2\pi i}{3}}\,,\0
\ee
we get
\be
[\hat V_{(i)}^{ab}r^*z]_{nj}&=&\delta^{ab}t_{nj}-\hat v^{ab}_{nj}-2i\delta^{ab}V_n^{(2)}(-2i)V_{j}^{(-1)}(2i)+\xi_3\alpha^{a-b}
 V_n^{(2)}(-\xi_3)V_{j}^{(-1)}(\xi_3)\0\\&&\!+
\left(\frac{\xi_3}{2}-i\delta^{ab}\right)V_n^{(2)}(0)V_{j}^{(-1)}(0)\0\\
&=&[\delta^{ab}r-\hat v^{ab}_{(i)}]_{nj}\,.
\ee

\noindent We can repeat the same procedure for the $Y(-i)$--insertion and prove that
\be
 \hat v_{(-i)}^{ab}&=&\delta^{ab}r^*- \hat V_{(-i)}^{ab}rz
\ee

\noindent These identities will be useful in order to simplify the midpoint product of two $gh$=0 states, as we will see later.

%%%%%%%%%%%%%%%%%%%%%%%%%%%%%%%%%%%%%%%%%%%%%%%%%%%%%%%
\subsection{$gh$=0 from $gh$=3: zero modes coefficients}
%%%%%%%%%%%%%%%%%%%%%%%%%%%%%%%%%%%%%%%%%%%%%%%%%%%%%%%

The operator $Y(z)$ has a non trivial kernel so it would seem that, by reflecting a state, some $gh$=0 information will get necessarily lost.
This would be the case indeed if we used only one insertion (say $Y(+i)$).
In order to appreciate this let us concentrate for a moment on the subalgebra of surface states. Let $\ket f$ be the surface state associated (modulo
$SL(2,C)$) to a function $f$, holomorphic on the unit semidisk of the $UHP$. Let then $\ket{\hat f_{(i)}}=Y(i)\ket f$ be the $gh$=3 'dual' of $\ket f$. 
Observing that
$$b_{-1}b_0b_1Y(i)\ket0=\ket0,$$  it follows that
$$f^{-1}\circ b_{-1}\;f^{-1}\circ b_0\;f^{-1}\circ b_1\ket{\hat f_{(i)}}=\ket f.$$
Now, for $i=-1,0,1$, we have $$f^{-1}\circ b_i\ket f=(b_i-F_{im}b_m^\dagger)\ket f=0, $$
where $F_{im}$ are the zero modes of the Neumann matrix of $\ket f$.
So we get
$$\prod_{i=-1,0,1}(b_i-F_{im}b_m^\dagger)\ket{\hat f_{(i)}}=\ket f.$$
Calling $\hat F$ and $F$ the Neumann matrices of $\ket{\hat f_{(i)}}$ and $\ket f$ respectively we then get the relation
\be
F_{nm}=\hat F_{nm}+\hat F_{n,-j}F_{jm},\quad\quad n,m\geq2,\quad j=-1,0,1
\ee
which is the same relation we get using the midpoint reflector, see below.
Notice that, without knowing in advance the zero mode components $F_{in}$, we cannot go back at $gh$=0. This is a consequence of the fact that
$Y$ has a non trivial kernel.\footnote{It is clear that, if we limit ourselves to surface states, this treatment is kind of overshooting since, for such states,
 starting from  the $gh$=3 Neumann function (\ref{s3}), one can adapt the methods of \cite{Fuchs:2004xj} to directly derive the surface--state
function $f(z)$. Knowing this
function, one can then write down the Neumann coefficients at $gh$=0. However our intent here is to establish a general algebraic procedure to go
back at $gh$=0, which can possibly be extended to more general squeezed states.}\\

\no That's the basic reason why the insertions we use are actually two (the holomorphic
and the anti--holomorphic midpoint). The simultaneous knowledge of these two $gh$=3 states will be enough to reconstruct the corresponding
$gh$=0 squeezed state. We would like now to  show how the $gh$=0 zero modes are related to the $gh$=3 Neumann matrices.
To this end we notice that, by using the midpoint identity (\ref{mid-id}),  the $gh$=0/$gh$=3 relations for the $Y(+i)$ insertion, (\ref{gh0/gh3}), can
be written as \footnote{We are now comparing two {\it bra's} so we get rid of the twist matrices in (\ref{gh0/gh3})}
\be
S-\hat S_{(i)}&=&\hat s_{(i)}zs,
\ee
while doing the same for the $Y(-i)$ insertion we get\footnote{If the $gh$=0 state $\ket S$ is twist invariant, then the result from the $Y(-i)$ insertion
is obtained by simple twist conjugation of the $Y(i)$ insertion. However, in general, the star product of two twist invariant states $\psi$ and $\phi$
is not twist invariant so
the two holomorphic/anti--holomorphic insertions are still needed to get back the star--product $\psi*\phi$ from $Y(\pm i)(\psi*\phi)$.}
\be
S-\hat S_{(-i)}&=&\hat s_{(-i)}zs.
\ee

\noindent Taking the difference of the two we end up with%\footnote{Note that this equation would not exist for a single insertion.}
\be\label{zerogen}
\hat S_{(i)}-\hat S_{(-i)}=(\hat s_{(-i)}-\hat s_{(i)})zs.
\ee
Our task is to solve this equation for $s$ ($gh$=0 zero mode contribution). A quick inspection, (\ref{gh0/gh3}), on the involved degrees of freedom shows that
this equation is indeed consistent since it is an equality between two (at most) rank 3 matrices. But we can be more explicit.
In case of wedge states (where exact reconstruction formulas are available at $gh$=3) it is easy to solve this equation for $s$. For using the known
reconstruction formulas, we have ($\k=0$ does not affect the bulk)
\be
(\hat S_{(i)}-\hat S_{(-i)})_{nm}&=&-\xi_N\left(V^{(2)}_n(-\xi_N)V^{(-1)}_m(\xi_N)+ V^{(2)}_n(\xi_N)V^{(-1)}_m(-\xi_N)\right)\0\\
(\hat s_{(-i)}-\hat s_{(i)})_{nj}&=&\xi_N\left(V^{(2)}_n(-\xi_N)V^{(-1)}_j(\xi_N)+V^{(2)}_n(0)V^{(-1)}_j(0)+ V^{(2)}_n(\xi_N)V^{(-1)}_j(-\xi_N)\right),\0
\ee
where, as usual, $\xi_N=\frac{4i}{N}$.
First, we notice that in order to be able to reproduce (\ref{zerogen}) we  take
\be
V^{(-1)}_i(0)\,z_{ij}\,s_{im}=V^{(-1)}_{-i}(0)\,s_{im}=0,
\ee
that is
\be
s_{-1,n}=-s_{1,n}.
\ee
In this way we recover the  fact that $gh$=0 wedge states are annihilated by $b_{1}+b_{-1}$.\\
Now it is easy to see that the only form $s_{in}$ can take, in order to satisfy (\ref{zerogen}), is
\be\label{zero-ansatz}
s_{in}=v^{(2)}_i(-\xi_N)V^{(-1)}_m(\xi_N)+ v^{(2)}_i(\xi_N)V^{(-1)}_m(-\xi_N),
\ee
where $v^{(2)}_i(\pm\xi_N)$ are 3--dimensional vectors to be determined by inserting this expression in (\ref{zerogen}). This gives the following
orthogonality constraints (which also yield, consistently with our previous assumption, $(b_{1}+b_{-1})\ket{N}=0$)
\be
V^{(-1)}_{-i}(\pm\xi_N)\,v^{(2)}_i(\mp \xi_N)&=&-1\\
V^{(-1)}_{-i}(\pm\xi_N)\,v^{(2)}_i(\pm \xi_N)&=&0\\
V^{(-1)}_{-i}(0)\,v^{(2)}_i(\pm \xi_N)&=&0.
\ee

\no These orthogonality relations completely constrain   the $v^{(2)}$'s to be

\be
 v^{(2)}_j(\pm\xi_N)=\frac14 \left(\frac{N^2}{4},\,\mp \frac{iN}{2},\,-\frac{N^2}{4}\right), \quad j=-1,0,1
\ee
One can easily check that, by plugging this in the ansatz (\ref{zero-ansatz}), we exactly reproduce the $s_{in}$ as defined by the $gh=0$ Neumann function,
(\ref{wedge-function}).
 We stress again that this has been possible because of the two complementary insertions of $Y$ in $\pm i$.\\

%%%%%%%%%%%%%%%%%%%%%%%%%%%%%%%%%%%%%%%%%%%%%%
\subsubsection{A check: $gh=0$ zero mode coefficients of surface states}
%%%%%%%%%%%%%%%%%%%%%%%%%%%%%%%%%%%%%%%%%%%%%%

Since the possibility of getting back the $gh$=0 zero modes starting from $gh$=3 data is a key--point of our strategy, we would like to elaborate
a little bit  on the actual solvability of the equation
$$
\hat S_{(i)}-\hat S_{(-i)}=(\hat s_{(-i)}-\hat s_{(i)})zs.
$$
By construction this equation can be solved for all $gh$=3 matrices coming from reflecting $gh$=0 ones. This is however an empty statement unless
we have a general independent principle to identify a $gh$=3 matrix
which is the result of a midpoint reflection. In order to appreciate the problem, let us define
\be
\Gamma_{nm}&\equiv&[\hat S_{(i)}-\hat S_{(-i)}]_{nm}\\
\gamma_{nj}&\equiv&[\hat s_{(i)}-\hat s_{(-i)}]_{nj}
\ee
We are supposed to know these quantities and we have to solve for $s_{in}$ the equation
\be\label{system}
\Gamma_{nm}=-\gamma_{n,-j}\,s_{jm}=-(\gamma_{n,-1}s_{1,m}+\gamma_{n,0}s_{0,m}+\gamma_{n,1}s_{-1,m})
\ee
This is an $\infty\times\infty$ set of equations for the $3\times\infty$ unknowns $s_{in}$ so, for a general $\Gamma$,
the system is over--constrained and
thus without solution. This is obvious: not all $gh$=3 matrices we could think of come from $gh$=0 by midpoint reflection.
We can try to solve this system by picking three different $n$'s and calling them $n_i$ with $i=-1,0,1$. This choice is not completely free,
but it should be done in such a way that
\be
\det_{ij}\gamma_{n_i,j}\neq 0.
\ee
Then, for any $m=2,...,\infty$, we can solve the $3\times3$ system
\be\label{system3}
\Gamma_{n_i,m}=-\gamma_{n_i,j}s_{-jm}
\ee
for $s_{jm}$ and get a definite answer. But, in general, the result will depend on the
particular choice of the three $n_i$'s.
Without extra inputs we don't know under which conditions the whole  $\infty\times\infty$ system (\ref{system}) will not be overdetermined.
%Looking explicitly at the structure of $\hat S_{(i)}$ and $\hat S_{(-i)}$, (\ref{gh0/gh3},\ref{gh0/gh3(-i)}), we see that, whatever the
%unknown $gh$=0 matrix $S$ is, the difference $\hat S_{(i)}-\hat S_{(-i)}$ is going to have maximal rank 3, because of the summation on the three zero
%modes indexes. In this case  there must be a solution.\\

We can check that the above system is not overdetermined  if we restrict to the subalgebra of surface states because in this case we have directly and
independently the Neumann functions at $gh$=3.
Here too, however, there is a subtlety in the case of  projectors because the two midpoints collapse to the
same point on the boundary thus giving $\hat S_{(i)}=\hat S_{(-i)}$. That's not a surprise: projectors are kind of singular objects which should correctly
be interpreted as a limit of a sequence of well behaved surface states. For wedge states and the sliver this just means that the $N\to\infty$ limit
should be taken at the very end.\\

Modulo this subtlety,  (\ref{system}) can be explicitly solved for any surface state. In this case we have an explicit expression for
$\gamma_{n_i j}$ and $\Gamma_{nm}$ from the Neumann function which is just the $bc$ propagator on the surface state geometry in the presence of the
midpoint $Y$--insertion. We tried with many different surface states given by almost random holomorphic functions on the disk and (excluding the case
of projectors like the sliver, the butterflies, etc...) we always found that, when $  \det_{ij}\gamma_{n_i j}\neq 0$,
the solution of (\ref{system3}) for $s_{jm}$
is {\it independent} of the random choice of the $n_{i}$'s (and actually coincides with the $gh$=0 Neumann function).
Having found the $s_{jm}$'s one can then plug them into (\ref{system3}) when
$\det_{ij}\gamma_{n_i j}=0$ and verify consistency.\\

The strongly constrained linear system we have found above is reminiscent of  the fact that surface states are known to satisfy
the Hirota equations for the
KP hierarchy \cite{Bonora:2002un, Boyarsky:2002yh}(or the corresponding hierarchy for the ghost sector, \cite{Fuchs:2004xj}), which implies that the whole Neumann matrix is known once a single row/column is.
This leaves us with the possibility that, perhaps, it is  only for surface states that the whole picture is consistent. We think that this point deserves
further investigation, which, however, would go beyond the scope of this paper; so we leave it as an interesting open  problem.\\

%%%%%%%%%%%%%%%%%%%%%%%%%%%%%%%%%%%%%%%%%%%%%%%%%%%%%%%%%%%%%%%%%%%%%
\subsection{$gh$=0 from $gh$=3: bulk reconstruction}
%%%%%%%%%%%%%%%%%%%%%%%%%%%%%%%%%%%%%%%%%%%%%%%%%%%%%%%%%%%%%%%%%%%%%

Our strategy is to use the midpoint reflector to relate $gh$=0 Neumann
matrices to $gh$=3 ones and hence to infer a spectral decomposition of
the former by knowing the one of the latter. The reflector gives the
$gh$=3 Neumann matrices in terms of the $gh$=0 ones, see (\ref{gh0/gh3'})
\be \hat S_{(i)}&=&S+(Sr^*-rz)\frac{1}{1-sr^*}s.
\ee
Because of the denominator (which arises from a geometric
resummation) this relation is very complicated; also, our logic is
to {\it derive} the $gh$=0 Neumann matrices by knowing the $gh$=3 ones
(whose reconstruction is transparent), so we solve it for $S$
($gh$=0 bulk) and get
\be
 S=\hat S_{(i)}-(\hat
S_{(i)}r^*-rz)s\label{gh03}
\ee
This equation is now easy to check
analytically using the known reconstruction formulas. In particular
for wedge states we can use the relations $(zs={\cal C}s)$ and $({\cal C}r{\cal C}=r^*)$
and rewrite (\ref{gh03}) as
\be S=\hat S_{(i)}-(\hat S_{(i)}-{\cal C})r^*s.
\ee
The reconstruction formulae for all the terms on
the right hand side are known and are given by
\be
\hat S_{(i)nl}=\int_{\k>\xi_{N}}d\k \frac{{\mathfrak t}_N(\k)}{2\sinh \frac{\pi
\k}2} V_n^{(2)}(-\k)V_l^{(-1)}(\k)=\left(\int_R-\oint_{\xi_N}-\frac 12
\oint_{0}\right)(...),\0
\ee
 \be
 (\hat S_i-{\cal C})_{nm}=\int_{\k>\xi_{N}}d\k
\frac{{\mathfrak t}_N(\k)-1}{2\sinh \frac{\pi \k}2}
V_n^{(2)}(-\k)V_m^{(-1)}(\k)\0 \ee \be
(r^*s)_{ml}=-\int_{\k'>2i}d\k' \frac{V_m^{(2)}(\k')}{2\sinh
\frac{\pi \k'}2} \frac
{N^2}{32}\Big(\k'(\k'+\xi_N)V_l^{(-1)}(\xi_N)+\k'(\k'-\xi_N)V_l^{(-1)}(-\xi_N)\Big),\0
\ee
where $R$ represents the real axis.\\
We recall that
\be
{\mathfrak t}_N(\k)=\frac{\sinh\frac{\pi\k(2-N)}{4}}{\sinh\frac{\pi\k N}{4}},\0
\ee
and we have also used
$$\sum_{i=-1}^1\,V_{-i}^{(-1)}(\k)v^{(2)}_i(\pm \xi_N)=\frac{N^2}{32}\k(\k\mp\xi_N).$$

\no To see the
cancelation of some of the terms we would like to write $\hat
S_{(i)nl}$ more explicitly as
\be
\hat S_{(i)nl}=\int_{R}d\k
\frac{{\mathfrak t}_N(\k)}{2\sinh \frac{\pi \k}2}
V_n^{(2)}(-\k)V_l^{(-1)}(\k)-\xi_{N}V_n^{(2)}(-\xi_N)V_l^{(-1)}(\xi_N)-\frac
12 \oint_{0}(...).\label{sinl}
\ee
 Now, to evaluate the product $$(\hat
S_i-{\cal C})_{nm}(r^*s)_{ml},$$ we deform the $\k'>2i$ contour to
$\k'>\xi_{N}$ and we  pick the residue at $\k'=2i$ on the way. Since ${\mathfrak t}_N(2i)=1$, this residue is
actually vanishing.\\
\no Then we can write
\be
\Big((\hat S_i-{\cal
C})r^*s\Big)_{nl}&=&-\int_{\k>\xi_N}d\k
\frac{({\mathfrak t}_N(\k)-1)V_n^{(2)}(\k')}{2\sinh \frac{\pi
\k'}2}\label{gh03-1}\\&\times& \frac
{N^2}{32}\Big(\k(\k+\xi_N)V_l^{(-1)}(\xi_N)+\k(\k-\xi_N)V_l^{(-1)}(-\xi_N)\Big).\0
\ee

Now we close the contour at infinity, above the line $\k=\xi_N$, so that the value of
this integral will be given by the sum of all the residues at
$\k=2im,~~m=1,2...$ and at $\k=m\xi_N,~~ m=2,3,...$, where we
assumed $N>2$. For a reason which will become clear soon we start
the summation over the residues at $\k=m\xi_N$ from $m=1$ and then
subtract the $m=1$ contribution. Then we  obtain
\be
\Big(\hat
S_{(i)}-{\cal C})r^*s\Big)_{nl} =&&\oint\frac{dz}{2\pi
i}\frac{1}{z^{n-1}2\xi_N^2}\Big\{\Big(-\frac2{z^3}+\frac2{z}-\frac{\xi_N}{z^2}\Big)V_{l}^{(-1)}(\xi_N)\0\\
&&~~~~~~~~~~~~~~~~~~~~~+\Big(-\frac2{z^3}+\frac2{z}+\frac{\xi_N}{z^2}\Big)V_{l}^{(-1)}(-\xi_N)\Big\}\0\\
&&+\oint\frac{dz}{2\pi
i}\frac{\xi_N(1+z^2)^{-2}}{z^{n-1}}\Big\{\frac{e^{2\xi_N\tan^{-1}z}}{(-1+e^{\xi_N\tan^{-1}z})^3}V_{l}^{(-1)}(\xi_N)\0\\
&&~~~~~~~~~~~~~~~~~~~~~~~~+\frac{e^{\xi_N\tan^{-1}z}}{(-1+e^{\xi_N\tan^{-1}z})^3}V_{l}^{(-1)}(-\xi_N)\Big\}\0\\
&&-\xi_{N}V_n^{(2)}(-\xi_N)V_l^{(-1)}(\xi_N),\label{diff}
\ee
where
the last term is the contribution from $m=1$.  We also recall that
the $\frac 12 \oint_{0}(...)$ of (\ref{sinl}) does not affect the
bulk. Therefore, the right hand side of (\ref{gh03}) is given by
\be
\Big(\hat S_{i}-(\hat S_{i}-{\cal C})r^*s\Big)_{nl}
=&&-\oint\frac{dz}{2\pi
i}\frac{1}{z^{n-1}2\xi_N^2}\Big\{\Big(-\frac2{z^3}+\frac2{z}-\frac{\xi_N}{z^2}\Big)V_{l}^{(-1)}(\xi_N)\0\\
&&~~~~~~~~~~~~~~~~~~~~~+\Big(-\frac2{z^3}+\frac2{z}+\frac{\xi_N}{z^2}\Big)V_{l}^{(-1)}(-\xi_N)\Big\}\0\\
&&-\oint\frac{dz}{2\pi
i}\frac{\xi_N(1+z^2)^{-2}}{z^{n-1}}\Big\{\frac{e^{2\xi_N\tan^{-1}z}}{(-1+e^{\xi_N\tan^{-1}z})^3}V_{l}^{(-1)}(\xi_N)\0\\
&&~~~~~~~~~~~~~~~~~~~~~~~~+\frac{e^{\xi_N\tan^{-1}z}}{(-1+e^{\xi_N\tan^{-1}z})^3}V_{l}^{(-1)}(-\xi_N)\Big\}\0\\
&&+\int_{R}d\k \frac{{\mathfrak t}_N(\k)}{2\sinh \frac{\pi \k}2}
V_n^{(2)}(-\k)V_l^{(-1)}(\k). \ee
The first term vanishes except
for the first three rows. In this case the contribution from this
term is exactly canceled by the contribution from the second term.
Therefore, in the bulk calculation we can drop the first term and we
obtain \be \Big(\hat S_{i}-(\hat S_{i}-{\cal C})r^*s\Big)_{nl}
=&&-\oint\frac{dz}{2\pi
i}\frac{\xi_N(1+z^2)^{-2}}{z^{n-1}}\Big\{\frac{e^{2\xi_N\tan^{-1}z}}{(-1+e^{\xi_N\tan^{-1}z})^3}V_{l}^{(-1)}(\xi_N)\0\\
&&~~~~~~~~~~~~~~~~~~~~~~~~+\frac{e^{\xi_N\tan^{-1}z}}{(-1+e^{\xi_N\tan^{-1}z})^3}V_{l}^{(-1)}(-\xi_N)\Big\}\0\\
&&+\int_{R}d\k \frac{{\mathfrak t}_N(\k)}{2\sinh \frac{\pi
\k}2}V_n^{(2)}(-\k)V_l^{(-1)}(\k).\label{gh03-3}
\ee
 This formula is
meant to calculate only the bulk of the $gh=0$ wedge state. However,
it also gives the correct value of the first three rows, in which case
the second term is not contributing and the first term is exactly
the $s_{in}$ we have already obtained.

\no In total, the reconstruction of $gh$=0 wedge states is
given by
\be S_{Mn}=\int_R d\k\, \frac{{\mathfrak t}_N(\k)}{2\sinh\frac {\pi
\k}{2}}\,V_m^{(2)}(-\k)V_n^{(-1)}(\k) -
\sum_{\xi_N=\pm\frac{4i}{N}}\,\xi_N\,{\EuScript V}_M^{(2)}(-\xi_N)V_n^{(-1)}(\xi_N)\label{gh0rec}
\ee

\no The `exotic' h=2 vector ${\EuScript V}_M^{(2)}$ is explicitly given by (its explicit
form can be read from (\ref{gh03-3}))
 \be
{\EuScript V}_M^{(2)}(\xi_N)&=&\oint_0\frac{dz}{2\pi i}\,\frac1{z^{M-1}}F^{(2)}_{\xi_N}(z)\\
F^{(2)}_{\xi_N}(z)&=&\frac{1}{(1+z^2)^2}\left(\frac1{1-f_{N}(z)}\right)^3\,f_{N}(z)
\ee where $f_{N}(z)$ is the wedge mapping function
\be
f_{N}(z)=\exp(\xi_N\tan^{-1}z)=\left(\frac{1+iz}{1-iz}\right)^{\frac2N}
\ee
It should be noted that ${\EuScript V}^{(2)}_M$, when evaluated at $\xi_N=\pm 2i$, coincides, up to a normalization, with the discrete $v^{(2)}(\pm 2i)$ eigenvectors of $G$,
(\ref{V(2)}).\\

One can explicitly check that the above reconstruction formula exactly reproduces  the
Neumann coefficients of $gh$=0 wedges, (\ref{wedge-function}).

\no As an example we show that  a sample of the entries for $N=5$ coincide with the residues of (\ref{wedge-function})
\begin{eqnarray}
S_{-1,3}^{(N=5)} &=& -\sum_{\xi_5=\pm\frac{4i}{5}}\,\frac{4i}{5}\,{\EuScript V}_{-1}^{(2)}(-\xi_5)V_3^{(-1)}(\xi_5) = -\frac{7}{25} \0 \\
S_{0,2}^{(N=5)} &=& -\sum_{\xi_5=\pm\frac{4i}{5}}\,\frac{4i}{5}\,{\EuScript V}_{0}^{(2)}(-\xi_5)V_2^{(-1)}(\xi_5) = -\frac{14}{25} \0 \\
S_{1,3}^{(N=5)} &=& -\sum_{\xi_5=\pm\frac{4i}{5}}\,\frac{4i}{5}\,{\EuScript V}_{1}^{(2)}(-\xi_5)V_3^{(-1)}(\xi_5) = +\frac{7}{25} \0 \\
S_{2,2}^{(N=5)} &=& \int_{-\infty}^{\infty}   d\k \frac{{\hat {\mathfrak t}}^{(5)}
(\k)}{2 {\rm sinh} (\frac{\pi \k}{2})} V_2^{(2)}(-\k)V_2^{(-1)}(\k) \0\\&-&
\sum_{\xi_5=\pm\frac{4i}{5}}\,\xi_5\,{\EuScript V}_2^{(2)}(-\xi_5)V_2^{(-1)}(\xi_5)=-\frac{651}{3125}+\frac{448}{625} = \frac{1589}{3125}. \0
\end{eqnarray}
Even more amazingly, this reconstruction formula also works for the identity string field $N=1$ (for which we have $\Im(\xi_1)=4>\Im(2i)$),
where the same few entries still agrees with (\ref{wedge-function})
\begin{eqnarray}
S_{-1,3}^{(N=1)} &=& -\sum_{\xi_1=\pm\frac{4i}{1}}\,\frac{4i}{1}\,{\EuScript V}_{-1}^{(2)}(-\xi_1)V_3^{(-1)}(\xi_1) = 1 \0 \\
S_{0,2}^{(N=1)} &=& -\sum_{\xi_1=\pm\frac{4i}{1}}\,\frac{4i}{1}\,{\EuScript V}_{0}^{(2)}(-\xi_1)V_2^{(-1)}(\xi_1) = 2 \0 \\
S_{1,3}^{(N=1)} &=& -\sum_{\xi_1=\pm\frac{4i}{1}}\,\frac{4i}{1}\,{\EuScript V}_{1}^{(2)}(-\xi_1)V_3^{(-1)}(\xi_1) = -1 \0 \\
S_{2,2}^{(N=1)} &=& \int_{-\infty}^{\infty}   d\k \frac{{\hat {\mathfrak t}}^{(1)}
(\k)}{2 {\rm sinh} (\frac{\pi \k}{2})} V_2^{(2)}(-\k)V_2^{(-1)}(\k) \0\\&-&
\sum_{\xi_1=\pm\frac{4i}{1}}\,\xi_1\,{\EuScript V}_2^{(2)}(-\xi_1)V_2^{(-1)}(\xi_1)=1+0 = 1. \0
\end{eqnarray}
 This formula (as all the
reconstruction formulas in general) is non vanishing also for the
($ij$) block and the ($ni$) 3--column, these parts are however eliminated by the normal ordering on $\ket0$.\\

\no Perhaps with a slight abuse of language we call (\ref{gh0rec}) the spectral representation of the $gh$=0 wedges. With its help it is
possible to verify directly the nature of the difficulties met in App.C of II when trying to compute their eigenvalues.

%%%%%%%%%%%%%%%%%%%%%%%%%%%%%%%%%%
\subsection{Normalization}
%%%%%%%%%%%%%%%%%%%%%%%%%%%%%%%%%%
In the introduction we defined $gh$=0 and $gh$=3 wedges with no normalization in front (because the normalization would be the one point function of
$Y$ on the surface--state geometry).
However we have seen that (see section (4.1)) when a $gh$=0 squeezed state is reflected to a $gh$=3 one, a normalization is also produced. Since, for squeezed states, BRST invariance
implies that the overlap (see appendix A) $$<gh=3|gh=0>=1,$$ the appearance of this normalization would seem to ruin everything.
But looking closer at this normalization

\be
\N=\det(1-{\cal S}R)=\det_{3\times3}\left(1-sr\right)
\ee
it is easy to recognize that this expression is just the overlap of the dual $gh$=3 vacuum $\bra0Y(i)$(whose Neumann matrix is bulkless,
just the 3--column $r$)
with the wedge state $\ket n$ (the matter
part does not contribute in this case, it is purely ghost business). Now, we check in appendix A that, when universal regularization is used,
\cite{FKuni}, we have
\be
\N=\bra0Y(i)\ket n=\bra{\hat 2_i}\ket{n}=1
\ee
So we see that, in total, the reflector does not produce any normalization when it acts on a BRST $gh$=0 squeezed state.

\no This will be valid in general (see \cite{Aref'eva:2006pu, Fuchs:2005ej, Belov1, FKuni} for previous, related works on this):
{\it The total matter+ghost normalization which is produced by the midpoint star product is globally 1}.\\
\no In particular we could generalize our construction to $N$--strings midpoint--vertices, \cite{carlo-new}, with the expectation that

 $$\bra {\hat V_{(i)}{}_N}\ket I=\bra{\hat V_{(i)}{}_{N-1}}$$
will hold with no normalizations in front.

%%%%%%%%%%%%%%%%%%%%%%%%%%%%%%%%%%%%%%%%%%%%%%%%%%%%%%%%%%%%%%%%%%%%%%%%%%%%%%
\section{Midpoint multiplication of $gh$=0 squeezed states}
%%%%%%%%%%%%%%%%%%%%%%%%%%%%%%%%%%%%%%%%%%%%%%%%%%%%%%%%%%%%%%%%%%%%%%%%%%%%%%

In paper II we saw that {\it if} we were allowed to star multiply $gh$=3 matrices (which, upon twisting with the ${\cal C}$--matrix
 are reconstructed by the same upper path as the  twisted matrices defined by the vertex), all the matrices in the game would commute and
the wedge recursion relations would arise automatically, exactly as for the matter sector. Here we are going to show that, as far as the product
matrix is concerned, there is no difference in using $gh$=3 matrices  or $gh$=0 ones. This will be a consequence of the midpoint
identities and the $gh$=0/$gh$=3 relations which we just derived.

 \no In order to avoid heavy computations, we will limit ourselves to\footnote{The
case with two general squeezed states works in exactly the same way.}
\be
\bra {\hat V_{(i)}{}_3}\ket N\ket0=\bra{\widehat{N+1}_{(i)}}
\ee
Note that
we intentionally avoided writing any normalization constant: we are thus assuming that a non--universal matter sector (for example 26 free real bosons)
is coupled to make $c_{tot}=0$, ghosts alone cannot be consistent after all. In the absence of normalizations this means

\be
Y(i)(\ket N*\ket 0)=Y(i)\ket{N+1}
\ee
The explicit expressions for the midpoint--vertex are (see II)
\be
\bra{\hat V_{(i)}}&=&\bra {V_3}Y(i)=\bra{\hat0}\, e^{-c_n^r\,\hat V_{(i)}{}^{rs}_{nM}\,b_M^s}\\
\hat V_{(i)}^{rs}(z,w)&=&\left[\frac{\partial f_3^r (z)^2}{\partial f_3^s(w)}
\frac{1}{f_3^r(z)-f_3^s(w)}\left(\frac{f_3^s(w)}{f_3^r(z)}\right)^3-\frac{\delta^{rs}}{z-w}\right]\0\\
&=&\hat V^{rs}(z,w)-\frac{4i}{3}\left(e^{\frac{2\pi i}{3}(r-s)}\,f^{(2)}_{-\frac{4i}{3}}(z)f^{(-1)}_{\frac{4i}{3}}(w)+
\frac12f^{(2)}_{\k=0}(z)f^{(-1)}_{\k=0}(w)\right)\0\\
&=&\int_{\k>\xi_3}\,\frac{v_3^{rs}(\k)}{2\sinh\frac{\pi \k}{2}}f^{(2)}_{-\k}(z)f^{(-1)}_{\k}(w),\quad\text{only in the bulk}.
\ee
In passing from the 3rd to the 4th row we took the $\k$ path above $\Im(\k)=\xi_3$ and we restricted ourselves just to the bulk, so that we can ignore
the discrete spectrum and the  $\k=0$ contribution.\\

\no This, together with the previous properties extracted from the 2--vertex (and from the known properties of the  vertex)
is all we need. In particular we recall the midpoint identities
\be
\hat v_{(i)}^{ab}&=&\delta^{ab}r-\hat V_{(i)}^{ab}r^*z.\label{midpid}
\ee
By explicit use of the squeezed states formula, we get (we use a boldface notation to represent (ls) and (sl) matrices and a plain
one for their (ss) bulk)
\be
\bra {\hat V_3{}_{(i)}}\ket N\ket0=\N\, e^{-c_n\, \hat{\bf[S*0]}_{(i)}{}_{nM}\,b_M},
\ee
with
\be
\N=\det (1-\hat {\bf V}_{(i)}^{11}{\bf S}).
\ee
First,  since $$\hat{\bf V}_{(i)}^{11}=\hat{\bf S}_{(i)}|_{N=3},$$ this is just the ghost part of the overlap $\bra{\hat 3_{(i)}}\ket{N}$
which equals unity,
when the matter sector is added and universal regularization is used, see appendix A.
As claimed, the vertex does not produce normalizations on BRST invariant states. \\

\no Let us now turn to the matrix defining the product, which  is given by
\be
\hat {\bf [S*0]}_{(i)}=\hat{\bf V}_{(i)}^{11}+\hat{\bf V}_{(i)}^{12}{\bf S}\frac1{1-\hat{\bf V}_{(i)}^{11} {\bf S}}\hat{\bf V}_{(i)}^{21}.
\ee
As we did for the reflector, we block decompose the above expression and use the identities (\ref{midpid}) to get
\be
\hat{\bf [S*0]}_{(i)}=\left(\begin{matrix}0&\quad0\\r-\hat{(S*0)}_{(i)}r^*z &\quad\hat{(S*0)}_{(i)}\end{matrix}\right).
\ee
Notice that the vertex automatically implements the midpoint identity for the product state. The product--bulk is given by
\be
\hat{(S*0)}_{(i)}=\hat V_{(i)}^{11}+(\hat V^{12}_{(i)}S+\hat v^{12}_{(i)}s)\frac1{1-\hat V^{11}_{(i)} S-\hat v^{11}_{(i)}s}\hat V^{21}_{(i)}.
\ee

\no Here we see very clearly that, even if $gh=3$ matrices commute among themselves, when also $gh=0$ matrices enter the game, commutativity seems
to be lost because, in the product--bulk, we get contributions from the zero modes of the states we multiply. In the absence of a clear relation
between $gh=0$ matrices and $gh=3$ ones, we would not be able to proceed further.\\

But now we can use the results of section 4, in particular the midpoint identities (\ref{midpid}) and the $gh$=3/$gh$=0 relation
\be
S=\hat S_{(-i)}-(\hat S_{(-i)}r-r^*z)s,
\ee
which allow us to write
\be
\hat V^{12}_{(i)}S+\hat v^{12}_{(i)}s=\hat V^{12}_{(i)}\hat S_{(-i)}\,(1-rs),
\ee
and
\be
\frac1{1-\hat V^{11}_{(i)} S-\hat v^{11}_{(i)}s}=\frac{1}{1-rs}\,\frac1{1-\hat V^{11}_{(i)}\hat S_{(-i)}}.
\ee
Notice that the order of matrices matters in this case as $gh=0$ and $gh=3$ matrices
generically do not commute.
All in all the product Neumann matrix will just contain $gh=3$ ingredients.
\be
\hat{(S*0)}_{(i)}=\hat V_{(i)}^{11}+\hat V^{12}_{(i)}\hat S_{(-i)}\frac1{1-\hat V_{(i)}^{11} \hat S_{(-i)}}\hat V^{21}_{(i)}.
\ee
By twisting the matrices and using ${\cal C}S_{(-i)}{\cal C}=S_{(i)}$, we finally get
\be
{\cal C}\hat{(S*0)}_{(i)}=\hat X_{(i)}^{11}+\hat X^{12}_{(i)}\hat T_{(i)}\frac1{1-\hat X_{(i)}^{11} \hat T_{(i)}}\hat X^{21}_{(i)}.
\ee
This is quite nontrivial: {\it when two $gh$=0 matrices are midpoint multiplied, they are effectively represented in the star product by $gh$=3 ones.}\\

\no So, {\it as far as the Neumann matrix is concerned}, there is no difference in multiplying  $gh$=0 states or `pure bulk'
 $gh$=0 states whose Neumann matrix
is given by the $gh$=3 one (with the opposite $Y$--chirality, since it is a ket). \\

All we have said till now about the product  matrix is formally valid for any squeezed state.
But going back to our main interest on wedge states, we see that, if we write the above expression in terms of twisted matrices,
 only matrices reconstructed by the upper $\k$--path will enter the game, so their product
will be regular and hence they will all commute (because they are all reconstructed on the same basis and all the paths are homotopic, in the sense that they can be deformed into one another without crossing singularities). This means that we can substitute
matrices with eigenvalues, multiply the eigenvalues and then reconstruct the product matrix with the upper path common to all.
Then, from the knowledge of both  $\hat{(S*0)}_{(i)}$ and $\hat{(S*0)}_{(-i)}$, we can use the results of section 4 to go back to $gh=0$  and thus
finally show that
\be
\ket{n}*\ket{m}=\ket{n+m-1}.
\ee
We have thus completed our long journey. In particular
we have showed that, at the end of the day, everything works as easily as for the zero momentum matter sector where $K_1$--invariance directly
implies commuting Neumann matrices.
%%%%%%%%%%%%%%%%%%%%%%%%%%%%%%%%%%%%%%%%%%%%%%%%%
\section{Conclusions and discussion}
%%%%%%%%%%%%%%%%%%%%%%%%%%%%%%%%%%%%%%%%%%%%%%%%%%
Our initial task was to star--multiply two $gh$=0 wedge states in the oscillator formalism. Finally we can claim that we succeeded. To start with we found
an explicit regular oscillator definition of the operation
\be
Y(\pm i){\Big(}\ket n *\ket m{\Big)}=Y(\pm i)\ket{n+m-1}
\ee
which in oscillator language (taking the $bpz$ to send ket's into bra's) reads
\be\label{end}
\bra{\hat V^{(3)}_{(\pm i)}}\ket n \ket m=\bra{\widehat{n+m-1}_{(\pm i)}}
\ee
That's the only expression we can write in which
\begin{itemize}
\item All objects are BRST invariant
\item All objects are squeezed states
\item The vertex is cyclic in the  strings indices
\item All objects are annihilated by $K_1$
\end{itemize}

The first property is the most important one as it implies that, if we regulate everything with universal regularization, then the expression (\ref{end}) is
true {\it without any normalization in front}, so there is no conflict with the CFT method (in which normalizations never enter).\\

The vertex matrices are not twist invariant but obey a complex twist symmetry whose geometrical meaning is to exchange $Y(i)\leftrightarrow Y(-i)$.
We saw that this apparent pathology is not a problem at all, since the vertex matrices (in the bulk) are still completely commuting as their
reconstruction involves paths in the complex $\k$--plane which are all homotopic. Hence dangerous divergences from the poles in the imaginary
axis are avoided because they are never crossed (the path is in the ''$\k$--UHP'' for all matrices and it is in the same place for the product).
So the complex Neumann coefficients (and hence the lack of twist invariance) are just the result of shifting the path from the real line (`principal' part)
 to $\Im(\k)>\Im(\xi_N)$ (`principal' {\it plus} `residual' part).\\

\noindent While there is no problem of convergence when the paths are homotopic, non--homotopic paths will give rise to divergences,
 encoded in the explicit
appearance of complex delta functions (or, equivalently, in $f_N(-i)$ or $f_N'(\pm i)$ if we use the generating function method on the $z$--plane).
The generating function method is actually more trustable because, on the $z$--plane, universal regularization is just our usual branch--point
displacement. In all analyzed examples  the generating function method shows very clearly that
the divergences arise from just the {\it principal} poles at ($\k=-\xi_N,\,0,\,\xi_N$) and not from {\it secondary} poles at $\k=\pm n\xi_N$. Still
it would be desirable to understand this directly on the $\k$--plane.\\

More concretely, we related $gh$=3 and $gh$=0 Neumann matrices by means of the midpoint 2--string vertex which allowed us to uncover many interesting properties.
\begin{itemize}
\item All $gh$=3 reflected states obey a midpoint identity which relates the 3--column of their Neumann matrix to the bulk.

\item It is very important to have {\it two} complementary reflectors (corresponding to the insertion of $Y(\pm i)$) since this allows us to
{\it derive} the $gh$=0 zero mode reconstruction by using the known reconstructions at $gh$=3. The elusive zero mode contribution of the
 $gh$=0 Neumann matrices can be indeed derived by the combined use of the $Y$ insertion at the two midpoints $\z=\pm i$.
We checked this property for random surface states which are not projectors, finding perfect agreement.

\item Once the $gh=0$ zero modes are derived from the $gh=3$ matrices, the bulk part can also be easily reconstructed.
This clarifies why  we have simple commutation relations at work at $gh=3$, while this is not true at $gh=0$.

\item All the potential divergences arising from imaginary poles in the complex $\k$ plane are avoided because all the paths in the reconstruction
formulas can be deformed into one another without crossing (principal) poles, secondary poles should be ignored
and this is independently checked by computing the matrix products on the $z$-plane with universal regularization.
Secondary poles are nonetheless needed to go back to $gh$=0 where the
`residual' contributions (which include the complete zero mode sector) are  given by summing all of their residues on the imaginary axis.

%\item If two wedge matrices with a  reconstruction that is defined with an opposite path with respect to the real axis of the complex $\k$--plane
% were ever to enter the game, un--cancelable divergences would arise: it is
%amazing to see how the CFT construction of the Neumann matrices secretly knows about this.

\item The reflector produces a normalization, but this normalization is 1 in universal regularization (since it is just the overlap of
the dual vacuum with the wedge state). So, in total, the reflector takes a $gh$=0 BRST closed squeezed state with no normalization and gives back
a $gh$=3 BRST closed squeezed state with no normalization.
\end{itemize}

After exploration of the $gh=0$/$gh=3$ relation, we finally came to the conclusive point of this paper: in the midpoint product the violation of
commutativity at $gh=0$ is elegantly avoided and, in order to do the star product, one can use commuting $gh=3$ matrices instead of $gh=0$ ones. Thus everything works as in the
matter sector. After all the initial intuition of I was correct.

%%%%%%%%%%%%%%%%%%%%
\acknowledgments
%%%%%%%%%%%%%%%%%%%

C.M. would like to thank Ted Erler for interesting discussions.
L.B. would like to thank
the GGI Institute in Florence, where part of this research was carried out, for hospitality
and financial support.
\noindent C.M. and D.D.T. would like to thank SISSA for the kind hospitality during part of this research.
The work of D.D.T. was supported by the Korean Research Foundation Grant funded by the Korean Government with
grant number KRF 2009-0077423.
R.J.S.S is supported by CNPq-MCT-Brasil.

%%%%%%%%%%%%%%%%%%%%%%%%%%%%%%%%%%%%%%%%%%
\section*{Appendix}
%%%%%%%%%%%%%%%%%%%%%%%%%%%%%%%%%%%%%%%%
\appendix

\section{BRST invariant squeezed states}

It is  assumed, \cite{leclair, LeClair:1988sj}, that all  surface states are BRST invariant when the total central charge vanishes. We will also
assume that all surface states with insertions are also BRST invariant, if the insertions are given by BRST invariant primary operators.
Example of these operators are $Y(z)$, $c\partial c(z)$ (which are $gh$=3/$gh$=2 and purely ghost) or
$c(z)V^{(m)}(z)$ (with $V^{(m)}(z)$ a weight 1 primary matter field).

We concentrate from now on on $gh$=0 and $gh$=3  squeezed states, and in particular we zoom on the family
of wedge states. This enhances BRST invariance with an extra symmetry, which is $K_1=L_1+L_{-1}$. Wedge states are indeed the only surface states which
are annihilated by $K_1$. There can be other $gh$=0/$gh$=3 squeezed states which are annihilated by $K_1$ but which are not BRST invariant. These states
should be considered pathological; examples of these states are $gh$=0/3 `principal' wedges (whose bulk Neumann matrices are just
given by the continuous spectrum integrated along the real axis).
Such states can be used in intermediate computations because of their very simple structure,
but at the end one should add the `residual' contribution from the $\xi_N$ poles in order to restore BRST invariance.\\

\noindent How can we check BRST invariance? The most direct way is  to do it by brute force. As a first step one can also check $K_n=L_n+L_{-n}$
invariance,
which is in general anomalous for even $n$ and non--vanishing total central charge. These tests are of course possible,  but quite cumbersome.
However, while it is difficult to directly check BRST invariance, it is quite easy to {\it disprove} it, at least in the $gh$=0/3 sector.
Indeed the cohomology at $gh$=0/3 is one--dimensional, i.e. there is just one state, \cite{coho}. The cohomology representative
at $gh$=0 can be chosen to be the unit operator {\bf 1}. The representative at $gh$=3 can be chosen to be $Y(z)$. In terms of states,
the $gh$=0 representative is just the SL(2,R) vacuum $$\ket 0={\bf 1}(0)\ket 0,$$ while at $gh$=3 it is $$\ket{\hat 0 }=Y(0)\ket 0=c_{-1}c_0c_{1}\ket0.$$

Since we are focusing on the sector, it is convenient to normalize the space-time volume to $(2\pi)^D$ so that we have
$$\bra0Y(0)\ket0=\bra0c_{-1}c_0c_1\ket0=\frac V{(2\pi)^D}=1.$$

Now consider a pair of $gh$=0/$gh$=3 squeezed states in the total matter--ghost CFT
\be
\ket f&=&\exp\left (\frac12\,a^\dagger\cdot F^{(m)}\cdot a^\dagger\right) \exp\left(\sum_{N,m}\,c^\dagger_N\,F_{Nm}\,b^\dagger_m\right)\,\ket0=\ket0+(...)\0\\
\bra {\hat g}&=& \bra{\hat0}\,\exp\left (\frac12\,a \cdot G^{(m)}\cdot a\right) \exp\left(-\sum_{n,M}\,c_n\,G_{nM}\,b_M\right)=\bra{\hat 0}+(...)
\ee

If these two squeezed states are BRST invariant, it means that the $(...)$ are $Q$--exact (precisely because the two vacua exhaust the $gh$=0/3 cohomology).
Then we are lead to conclude that ($c_{tot}=0$, otherwise the notion of cohomology is not even defined)
\be
\bra {\hat g}\ket f=\bra0c_{-1}c_0c_1\ket0=1
\ee
That is: {\it BRST invariance implies unit scalar product for any pair of BRST invariant $gh$=0/3 squeezed states.}

Reversing this property:
{\it If $\ket f$ and $\bra {\hat g}$ are two $gh$=0/3 squeezed states and $\bra {\hat g}\ket f\neq1$, then at least one of the
two states  is not BRST invariant}.

We can thus use this property to show that some of the squeezed states which can be built using the $K_{1}$ basis are not Q--closed.
In fact, in the oscillator language the quantity $\bra {\hat g}\ket f$ is given by (bosons at the denominator, fermions at the numerator)
\be\label{scaprod}
\bra {\hat g}\ket f=\frac{\det (1-G_{sl}F_{ls})}{\det (1-G^{(m)}\cdot F^{(m)})^{\frac D2}}=\frac{\det (1-F_{ls}G_{sl})}{\det (1-G^{(m)}\cdot F^{(m)})^{\frac D2}}
\ee
If this quantity can be computed and turns out not to be 1, it means that (at least) one of the states is not BRST invariant. In practice
this ratio is very delicate, and when it is not 1 it is vanishing or diverging (for Neumann matrices of infinite rank).
One thus needs  a trustable regularization procedure to control this norm. This very non trivial regularization was put on a firm ground
by Fuchs and Kroyter, under the name of universal regularization, \cite{FKuni}. The idea is very simple: shrink the string field a little bit
with the operator $e^{sL_0}$  with $s\to 1^{-}$ in order to detach it from the midpoint while computing the overlaps. We actually used this regularization when computing
the product of infinite matrices on the $z$--plane in II.

In this particular case this regularization consists in regulating the (matter and ghost) matrices in the following universal (the same for matter
and ghosts) way
\be
F^{(m)}_{nm}&\to& s^{n+m}\,F^{(m)}_{nm}\\
G^{(m)}_{nm}&\to& s^{n+m}\,G^{(m)}_{nm}\\
F_{Nm}&\to&s^{N+m}F_{Nm}\\
G_{nM}&\to&s^{n+M}G_{nM},
\ee
where $s\to 1^{-}.$

\no For large $(nm)$ this regularization has the effect of truncating with an exponential cutoff, which is stronger than any other
power--law divergence which one can encounter. So, as far as $s<1$, the Neumann matrices will be effectively truncated to a finite level.
Let us give the relation between the $s$ regulator and level truncation. Using wedge states we empirically found
\be
1-s\sim \frac2L.
\ee

This relation should be understood in the following way:
\begin{itemize}
\item Pick an $s<1$ and compute numerically (\ref{scaprod}) by  using the regularized matrices. The numerical computation can only be done
with finite size matrices, so one has to  truncate all the matrices  to a level $L$ (the same for matter and ghost). The result will be finite.
\item One increases the level by keeping the same $s$, the exp-cutoff given by $s$ will assure convergence to a finite value as $L\to \infty$. The
limiting value will be almost given by a finite level $L\sim \frac{2}{1-s}$, after this level corrections will become very small.
\item Pick another $s$ which is closer to 1 than the previous one. Repeat the above procedure.
\item The closer is $s$ to 1, the higher should be the level in order to see a convergence pattern.
\end{itemize}
It is clear that one would like to be able to compute (\ref{scaprod})  analytically in the parameter $s$: this is certainly doable by generalizing
the techniques of \cite{FKuni} (which deal with bosonic ghosts) to our case. In any case, even within a numerical approach, this
procedure gives very unambiguous results which mark a sharp distinction between BRST invariant squeezed states and the others.
%%%%%%%%%%%%%%%%%%%%%%%%%%%%%%%%%%%%%%%%%%%%%
\subsection{BRST invariant $gh$=0/3 wedges}
%%%%%%%%%%%%%%%%%%%%%%%%%%%%%%%%%%%%%%%%%%%%%
Wedge states are given by ($gh$=0, matter plus ghosts)

\be
\ket {n}=e^{\frac12 a_p^\dagger\cdot M^{(n)}_{pq}\cdot a_q^\dagger}\, e^{c_N^\dagger\,S^{(n)}_{Nm}\,b_m^\dagger}\ket0
\ee

with the  defining matrices given by
\be
M^{(n)}_{pq}&=&\frac{1}{\sqrt{p\,q}}\oint_0 \frac{dz}{2\pi i}\oint_0 \frac{dw}{2\pi i} \frac1{z^{p}}\frac1{w^{q}}\left[\frac{f'_n(z)f'_n(w)}{f_n(z)-f_n(w)}-\frac 1{(z-w)^2}\right]\\
S^{(n)}_{Nm}&=&\oint_0 \frac{dz}{2\pi i}\oint_0 \frac{dw}{2\pi i} \frac1{z^{N-1}}\frac1{w^{m+2}}
    \left[\frac{f_n'(z)^2}{f_n'(w)}\frac{1}{f_n(z)-f_n(w)}\left(\frac{f_n(w)-f_n(0)}{f_n(z)-f_n(0)}\right)^3-\frac{w^3}{z^3(z-w)}\right]\0
\ee

We now want to consider $gh$=3 wedges. There is an infinite family of them, as they differ in the location where one inserts the operator $Y(z)$.
Out of the infinite  places where one can insert $Y$, it is only the midpoint which is consistent with $K_1$ invariance. If we want to
have a single squeezed state  we have two choices, $z=\pm i$

\be
\bra {\hat n_{(+i)}}&=&\bra {n}Y(+i)=\bra{\hat0}e^{\frac12 a_p\cdot M^{(n)}_{pq}\cdot a_q}\, e^{-c_m\,\hat S^{(n)}_{(i)}{}_{mN}\,b_N}\\
\bra {\hat n_{(-i)}}&=&\bra {n}Y(-i)=\bra{\hat0}e^{\frac12 a_p\cdot M^{(n)}_{pq}\cdot a_q}\, e^{-c_m\,\hat S^{(n)}_{(-i)}{}_{mN}\,b_N}
\ee
The only change wrt $gh=0$ is in the ghost Neumann matrices, (\ref{Si}, \ref{S-i}). \\

\no To give an example, we plot in  figure 1 the numerical evaluation of  $\bra{{\hat n}_{(i)}}\ket n$ for $n=3$ using (\ref{scaprod}).
\vspace{1cm}

\FIGURE{\epsfig{file=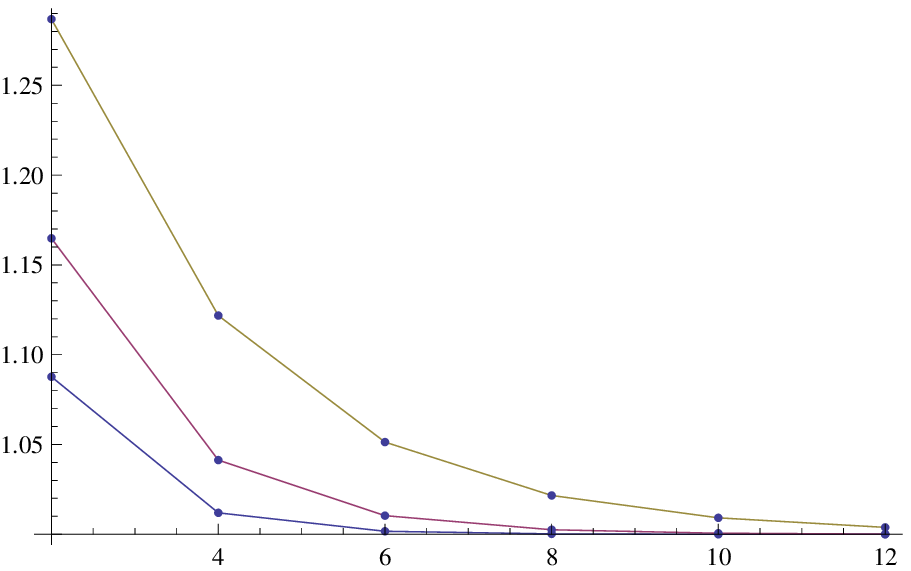,width=5cm} \epsfig{file=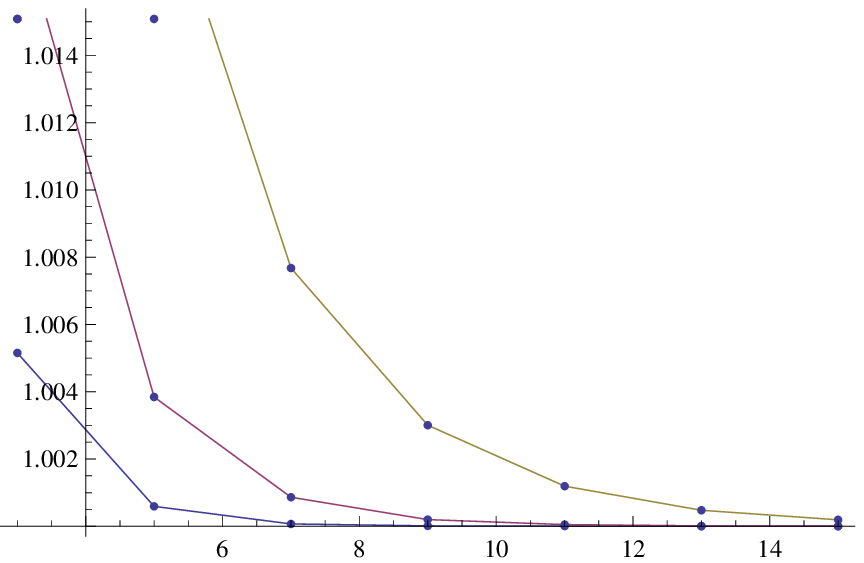,width=5cm}
        \caption { Evaluation of the overlap $\bra{{\hat 3}_{(i)}}\ket3$ for  even (on the left) and  odd (on the right)  levels.
The horizontal axis is the level (same for the matter and ghosts), while the vertical axis is
$\bra{{\hat 3}_{(i)}}\ket3$. Three values of the universal regulator $s$ are shown, $s=0.6,\;0.7,\;0.8$: $s=0.6$ corresponds to the set of points which (level by level) are closer
to the value $\bra{{\hat 3}_{(i)}}\ket3=1$, while $s=0.8$ corresponds to the most
distant set. Notice that, as $s$ gets closer to 1,  the level
should be increased in order to have a good convergence
to the expected unit value and that a reliable level is given by $L\sim2/(1-s)$.}}

\no It is easy to check that all the other overlaps with different wedges confirm that
\be
 \bra{\hat n_{(\pm i)}}\ket m=1,\quad\quad \forall\, n,m\geq1, \quad \forall\, s<1.
\ee
It is a quite encouraging result, which has to do with the self--consistency of the oscillator formalism with fermionic ghosts. More on this
is currently under investigation, \cite{carlo-new}.\\
As a counter--example, it is easy to check that, using the `principal' $gh$=3 wedges, instead of the BRST invariant ones, the norm (for fixed $s<1$)
will still converge to a finite value (although not the same for even and odd levels), but,
as $s$ approaches 1 the limit is more and more distant from unity : this
reveals a BRST breakdown. See figure 2 for an example.

\smallskip
\FIGURE{\epsfig{file= 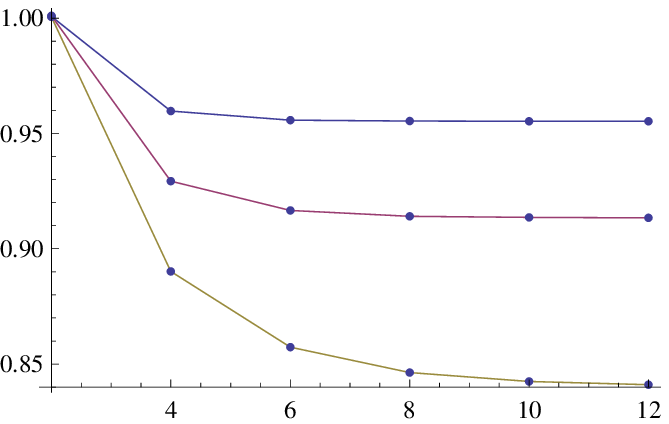,width=5cm} \epsfig{file= 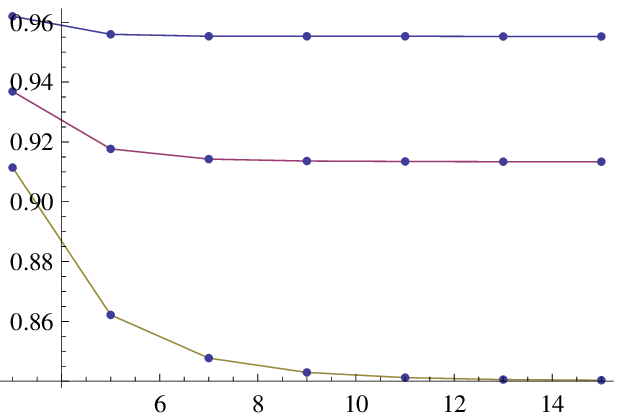,width=5cm}
        \caption {Evaluation of the overlap $\bra{{\hat 3}}\ket3$ for  even (on the left) and odd (on the right)  levels, using principal
wedges instead of $Y(\pm i)$--inserted one as $gh$=3 duals.
The three different sets of points are given by $s=0.6,\;0.7,\;0.8$: $s=0.6$ corresponds to the highest set of points in the plot,
while $s=0.8$ corresponds to the lowest set.
Notice that there is a convergence pattern for any value of $s<1$, but the limit gets far from unity as $s$ approaches 1.}}

%%%%%%%%%%%%%%%%%%%%%%%%%%%%%%%%%%%%%%%%%%%%%%%%%%%%%%%%%%%%%
\section{Irrelevance of secondary poles}
%%%%%%%%%%%%%%%%%%%%%%%%%%%%%%%%%%%%%%%%%%%%%%%%%%%%%%%%%%%%%

In this appendix we would like to show that secondary poles should not be considered when the product of two $gh=3$ matrices is performed. The
way we are going to show this is via an independent computation. We take as an example the proof of the midpoint identity
$$
\hat S_{(i)}r^*z=r-\hat s_{(i)},
$$
similar computations can be performed for all other products of $gh=3$ matrices which are considered in the main text.\\

\noindent First of all, let us trace  the point where secondary poles would give a contribution in the derivation of the above equation using the reconstruction
formulas. We have
\be
[\hat S_{(i)}r^*z]_{nj}=-\int_{\Im(\k')>2}\,d\k'\int_{\Im(\k)>\xi_N}\,d\k\,
\frac{{\mathfrak t}_N(\k)}{2\sinh\frac {\pi \k'}{2}}\,V_n^{(2)}(-\k)V_{j}^{(-1)}(\k')\delta(\k-\k')
\ee

In the main text we just assumed that, given $\xi_N<2i$, we could always take the $\k$--path to $\k>2i$, because there is no pole at $\k=2i$. However,
in the $\k$--UHP, the wedge eigenvalue ${\mathfrak t}_N(\k)$ have poles  at $\k=n\xi_N$ for all positive integers $n$'s. We call the $n=1$ pole a $principal$ pole,
while we call the others $secondary$ poles. The apparent problem here is that, starting from $N>4$, some secondary poles will be below $\Im(\k)=2$. So
they should be picked up when we shift the $\k$ path from $\k>\xi_N$ to $\k>2i$.\\

Taking into account the contribution from secondary poles we would have ended with
\be
[\hat S_{(i)}r^*z]_{nj}=[r-\hat s_{(i)}]_{nj}-4i\xi_N\sum_{k=2}^{n^*}\sinh\frac{\pi k\xi_N}{2}\delta(k\xi_N-2i)\,V_n^{(2)}(-k\xi_N)V_j^{(-1)}(2i),
\ee
where $n^*$ is the maximal integer for which $\frac{4}{N}n^*<2$. Notice the naked complex delta contribution which denounces an anomalous
divergent part. {\it In the main text we have assumed that one need not consider this contribution}.\\

We want to show here that what we did  is correct and it is what is implied by universal regularization. The way we are going to
show it is via an independent computation which, instead of using the power of reconstruction formulas, explicitly performs the matrix product via
generating functions defined on the ''Fourier transform'' of the $\k$ plane, the original $z$--plane (that is by the method extensively used in II). In doing this we explicitly consider the
`principal' + `residual' decomposition
\be
\hat S_{(i)}{}_{nm}&=&\oint_0 \frac{dz}{2\pi i}\oint_0 \frac{dw}{2\pi i} \frac1{z^{n-1}}\frac1{w^{m+2}}
{\Big (} \hat S(z,w)-\xi_N f^{(2)}_{-\xi_N}(z)\,f^{(-1)}_{\xi_N}(w){\Big )}\0\\
&=&\hat S_{nm}-\xi_N V^{(2)}_n(-\xi_N)\,V^{(-1)}_m(\xi_N),
\ee
and

\be
r^*_{mj}&=&\oint_0 \frac{dz}{2\pi i}\oint_0 \frac{dw}{2\pi i} \frac1{z^{m-1}}\frac1{w^{j+2}}
{\Big (}  t(z,w)+2i f^{(2)}_{2i}(z)\,f^{(-1)}_{-2i}(w)+i f^{(2)}_{0}(z)\,f^{(-1)}_{0}(w){\Big )}\0\\
&=&t_{mj}+2i V^{(2)}_m(2i)V^{(-1)}_j(-2i)+i V^{(2)}_m(0)V^{(-1)}_j(0).
\ee
The terms which are going to contribute to our problem are
$$
\sum_{m\geq2}\,{\Big(}-\xi_N V^{(2)}_n(-\xi_N)\,V^{(-1)}_m(\xi_N){\Big)}
{\Big(}2i V^{(2)}_m(2i)V^{(-1)}_j(-2i){\Big)}
$$
and
$$
\sum_{m\geq2}\,\hat S_{nm}{\Big(}2i V^{(2)}_m(2i)V^{(-1)}_j(-2i){\Big)}
$$
Let us begin by computing the inner product
$$
V^{(-1)}_m(\xi_N)V^{(2)}_m(2i).
$$
Using the continuous basis we get the formal expression
\be
V^{(-1)}_m(\xi_N)V^{(2)}_m(2i)=2\sinh\frac{\pi\xi_N}{2}\,\delta(\xi_N-2i).
\ee
Since $\xi_N\neq2i$, one would be tempted to take $\delta(\xi_N-2i)=0$, this is however not consistent with the explicit result we get by computing
the same quantity on the $z$ plane\footnote{Also, from general properties of analytic continuation, $\delta(ix)$ is a formal divergent quantity, for $x\in R$.}
\be
V^{(-1)}_m(\xi_N)V^{(2)}_m(2i)=\oint_0 \frac{dz}{2\pi i}\oint_0 \frac{dw}{2\pi i} \frac1{z^{m+2}}\frac1{w^{m-1}}\,
\frac{1+z^2}{(1+w^2)^2}\,f_N(z)\,\left(\frac{1+iw}{1-iw}\right),
\ee
where
$$
f_N(z)=\left(\frac{1+iz}{1-iz}\right)^{\frac2N}.
$$
As we saw in II, in order to sum up the geometric series, we have to regulate the wedge--function $\left(\frac{1+iz}{1-iz}\right)^{\frac2N}$ by
 pushing the branch points away from $z=\pm i$. As usual we do this by
$$
f_N(z)\to f_N^{(K)}(z)\equiv f_N(z/K).
$$
Then, as argued in II (there are no poles at $z=w$), in the integration,  we can choose the ordering
$$
\frac1K<\frac1z<w<1<\frac1w<z<K,
$$
and, by performing the integral around $w$, we end up with
\be
V^{(-1)}_m(\xi_N)V^{(2)}_m(2i)=\oint_i\frac{dz}{2\pi i}\frac1{(z-i)^2}f_N^{(K)}(z)=\frac{d}{dz}f_N^{(K)}(z){\Big |}_{z=i},
\ee
which is divergent. So, calling this divergence $\gamma$ we can make the association,
 $$\gamma\equiv \frac{d}{dz}f_N^{(K)}(z){\Big |}_{z=i} \longleftrightarrow 2\, \sinh\frac{\pi\xi_N}{2}\,\delta(\xi_N-2i).$$ Finally we have
\be
-\xi_N V^{(2)}_n(-\xi_N)\,V^{(-1)}_m(\xi_N)
2i V^{(2)}_m(2i)V^{(-1)}_j(-2i)=-2i\xi_N\;\gamma\; V^{(2)}_n(-\xi_N)V^{(-1)}_j(-2i).\label{div1}
\ee
This anomalous contribution needs to be compensated (and will) by the other term we consider,
$$
\sum_{m\geq2}\,\hat S_{nm}{\Big(}2i V^{(2)}_m(2i)V^{(-1)}_j(-2i){\Big)}.
$$
Here we need to compute
$$
\hat S_{nm}V^{(2)}_m(2i).
$$
Again, we will first compute this quantity on the un--regularized $\k$--plane (that is with the techniques of reconstruction formulas). This gives
\be
\hat S_{nm} V^{(2)}_m(2i)&=&{\mathfrak t}_N(2i)V_n^{(2)}(-2i)+2\xi_N\sinh\frac{\pi\xi_N}{2}\,V^{(2)}_n(-\xi_N)\,\delta(\xi_N-2i)\0\\
&+&2\xi_N\sum_{k=2}^{n^*}\sinh\frac{\pi k\xi_N}{2}\,V^{(2)}_n(-k\xi_N)\,\delta(k\xi_N-2i)\0
\ee
this expression is what one gets by taking the $\k$ path of $S_{nm}$ from the real axis to $\Im(\k)=2$, and accordingly picking up the
residues along the way. Notice the contribution from the principal pole at $\k=\xi_N$ (principal divergence), as well
as from the secondary ones at $\k=k\xi_N$ (secondary divergences). The divergence given by the principal pole exactly cancels with the previous
term, but secondary poles cannot be canceled against anything else.\\

\noindent We now show that, doing the same computation on the $regularized$ $z$--plane
(that is point--splitting plus contour integral techniques of section 2 of II), there will  be no contribution at all from secondary poles.
Explicitly (we do not write down the radial ordering term $\frac1{z-w}$ in $\hat S_{nm}$, because we are just interested in extracting midpoint singularities)
\be
\hat S_{nm} V^{(2)}_m(2i){\Big|}_{sing}&=&\oint_0 \frac{dz}{2\pi i}\oint_0 \frac{dw}{2\pi i}\oint_0 \frac{dx}{2\pi i}\frac1{z^{n-1}}\frac1{w^{m+2}}\frac1{x^{m-1}}\0\\
&&\frac{\xi_N}{2}\,\frac{1+w^2}{(1+z^2)^2}\,\frac{f_N(z)+f_N(w)}{f_N(z)-f_N(w)}\,\frac{1}{(1+x^2)^2}\left(\frac{1+ix}{1-ix}\right).
\ee
Again, in order to perform the sum over $m$, we regulate the wedge function
$$
f_N(w)\to f_N^{(K)}(w)\equiv f_N(w/K)
$$
and take
$$
\frac1K<\frac1w<x<1<\frac1x<w<K.
$$
We can now safely integrate over $x$ and get
\be
&=&\frac{\xi_N}{2}\oint_0 \frac{dz}{2\pi i}\frac1{z^{n-1}}\frac{1}{(1+z^2)^2}
\;\oint \frac{dw}{2\pi i}\frac1{(w-i)^2}\frac{f_N(z)+f^{(K)}_N(w)}{f_N(z)-f^{(K)}_N(w)}\0\\
&=&\frac{\xi_N}{2}\oint_0 \frac{dz}{2\pi i}\frac1{z^{n-1}}\frac{1}{(1+z^2)^2}\;
\frac{d}{dw}\left(\frac{f_N(z)+f^{(K)}_N(w)}{f_N(z)-f^{(K)}_N(w)}\right)_{w=i}\0\\
&=&\frac{\xi_N}{2}\oint_0 \frac{dz}{2\pi i}\frac1{z^{n-1}}\frac{1}{(1+z^2)^2}\;
\frac{2}{f_N(z)}\,\frac{d}{dw}f_N^{(K)}(w){\Big |}_{w=i}\0\\
&=&\xi_N\,\gamma\,V_n^{(2)}(-\xi_N)=2\xi_N\sinh\frac{\pi\xi_N}{2}\,V^{(2)}_n(\xi_N)\,\delta(\xi_N-2i)=\hat S_{nm} V^{(2)}_m(2i){\Big|}_{sing}.
\ee
In integrating over $w$ we just considered the pole at $w=i$, because it is the only one which is within the integration contour.\\

As claimed, the regularized computation on the $z$--plane (which is of course more trustable than the un--regularized computation on the $\k$ plane)
clearly shows that secondary poles should not be considered. A clear $\k$--plane argument for disregarding them should  emerge by
the use of universal regularization, \cite{FKuni}, (whose basic effect, on the $z$--plane, is just our branch--points displacement).
We leave this quite technical point for future investigations.\\

\no Similar checks can be done in all the other computations in the core of the paper: {\it when two $gh=3$ (wedge) matrices are multiplied, only
the poles at $\k=-\xi_N,0,\xi_N$ have to be considered and the homotopy class of the paths in the game will be always understood to be the one
implied by these principal poles.}

\end{document}